\documentclass[journal]{IEEEtran}

\usepackage{hyperref}

\usepackage{amsmath, amsfonts, amssymb}
\usepackage{algorithm, algorithmic}
\usepackage{graphicx, comment}
\usepackage{color,subfigure,cite,times}

\usepackage{url}
\usepackage{epsfig}
\usepackage{graphics,xspace}
\usepackage{multirow}
\usepackage{multicol}

\def\GF{\mathop{\sf GF}\limits}

\def\tran{\mathrm{T}}

\def\Z{{\mathbf{Z}}}

\def\x{{\mathbf{x}}}

\def\0{{\mathbf{0}}}
\def\1{{\mathbf{1}}}
\def\RR{{\mathbb{R}}}
\def\NN{{\mathbb{N}}}

\def\EE{{\mathsf{E}}}

\def\lmat{\left(\begin{matrix}}
\def\rmat{\end{matrix}\right)}
\def\eqref#1{(\ref{#1})}

\newtheorem{definition}{Definition}
\newtheorem{lemma}{Lemma}

\newtheorem{proposition}{Proposition}

\def\BibTeX{{\rm B\kern-.05em{\sc i\kern-.025em b}\kern-.08em
    T\kern-.1667em\lower.7ex\hbox{E}\kern-.125emX}}

\usepackage{setspace}
\usepackage{framed}

\newcommand{\be}{\begin{equation}}
\newcommand{\ee}{\end{equation}}

\definecolor{BrightBlue}{rgb}{0,0,1}

\def\av{{\mathbf{a}(t)}}
\newcommand{\avm}[1]{\ensuremath{a_{#1}(t)}}
\def\sv{{\mathbf{x}(t)}}
\newcommand{\svm}[1]{\ensuremath{\x_{#1}(t)}}
\newcommand{\ve}[1]{\mathbf{#1}}
\newcommand{\avg}[1]{\overline{#1}}

\def\buf{{\mathsf{Buffer}}}

\def\pkt{{\ensuremath{W}}}
\newcommand{\rcvpkt}[1]{{\ensuremath{W_{\text{rcvd},#1}}}}

\def\im{{\mathcal{A}}} 

\def\SCH{{\ensuremath{\mathsf{SCH}}}}

\def\cq{{\mathsf{cq}}}
\def\bigcq{{\mathsf{CQ}}}
\def\stdef{\stackrel{\Delta}{=}}
\def\vp{{\vec{p}}}
\def\rcpt{{\mathsf{rcpt}}}

\def\tran{{\mathsf{T}}}

\def\qinter{{\ensuremath{\ve{q}^{\text{inter}}}}}
\def\Qinter{{\ensuremath{\ve{Q}^{\text{inter}}}}}
\newcommand{\qkinter}[1]{q_{#1}^{\text{inter}}}
\newcommand{\Qkinter}[1]{Q_{#1}^{\text{inter}}}
\newcommand{\Nna}[1]{\ensuremath{N_{\mathsf{NA},#1}}}
\newcommand{\bin}[2]{\ensuremath{\beta_{#1,#2}^{\text{in}}}}
\newcommand{\bine}[2]{\ensuremath{\beta_{#1,#2}^{\text{in}}}}
\newcommand{\bout}[2]{\ensuremath{\beta_{#1,#2}^{\text{out}}}}
\def\Bin{{\ensuremath{\mathcal{B}^\text{in}}}}
\def\Bout{{\ensuremath{\mathcal{B}^\text{out}}}}
\def\Bini{{\ensuremath{\mathcal{B}^{\text{in},(i)}}}}
\def\Bouti{{\ensuremath{\mathcal{B}^{\text{out},(i)}}}}

\DeclareMathOperator*{\argmax}{arg\,max}

\def\eqdef{{\stackrel{\triangle}{=}}}

\newcommand{\SAin}[1]{\mathcal{I}_{#1}}
\newcommand{\SAout}[1]{\mathcal{O}_{#1}}

\newcount{\myi}
\newcommand{\Repeat}[2]{%
    \myi=0
    \loop
        \ifnum\myi<#2
        #1
        \advance\myi by 1
    \repeat
}

\definecolor{BrickRed}{rgb}{0,0,0}
\newcommand{\change}[1]{\textcolor{BrickRed}{#1}}
\newcommand{\ccwchange}[1]{{#1}}

\begin{document}

\title{Robust And Optimal Opportunistic Scheduling For Downlink 2-Flow Network Coding With Varying Channel Quality and Rate Adaptation (New Simulation Figures)}

\author{\IEEEauthorblockN{Wei-Cheng Kuo, Chih-Chun Wang}\thanks{This work was supported in parts by NSF grants ECCS-1407603, CCF-0845968 and CCF-1422997. Part of the results was presented in 2014 INFOCOM.}
\IEEEauthorblockA{\{wkuo, chihw\}@purdue.edu\\
School of Electrical and Computer Engineering, Purdue University, USA}
}

\maketitle

\begin{abstract}
This paper considers the downlink traffic from a base station to two different clients. When assuming infinite backlog, it is known that {\em inter-session} network coding (INC) can significantly increase the throughput. However, the corresponding scheduling solution (when assuming dynamic arrivals instead and requiring bounded delay) is still nascent.

For the 2-flow downlink scenario, we propose the first opportunistic INC + scheduling solution that is provably optimal for time-varying channels, i.e., the corresponding stability region matches the optimal Shannon capacity.
Specifically, we first introduce a new {\em binary INC} operation, which is distinctly different from the traditional wisdom of XORing two overheard packets. We then develop a {\em queue-length-based} scheduling scheme and prove that it, with the help of the new INC operation, achieves the optimal stability region with time-varying channel quality. The proposed algorithm is later generalized to include the capability of rate adaptation. Simulation results show that it again achieves the optimal throughput with rate adaptation.
A
byproduct of our results is a scheduling scheme for stochastic
processing networks (SPNs) with {\em random departure}, which relaxes the assumption of {\em deterministic
departure} in the existing results.
\end{abstract}

\section{Introduction\label{sec:introduction}}
Since 2000, network coding (NC) has emerged as a promising technique in communication networks. 
\cite{LiYeung03} shows that linear intra-session NC achieves the min-cut/max-flow capacity of single-session multi-cast networks. The natural connection between intra-session NC and the {\em maximum flow} allows the use of  back-pressure (BP) algorithms to stabilize intra-session NC traffic, see \cite{ho2009dynamic} and the references therein.

However, when there are multiple coexisting sessions, the benefits of {\em inter-session} NC (INC) are not fully utilized \cite{KhreishahWangShroff10,WangShroff10}. The COPE architecture \cite{KattiRahulHuKatabiMedardCrowcroft06} demonstrated that a simple INC scheme can provide 40\%--200\% throughput improvement in a testbed environment. Several analytical attempts have been made to characterize the INC capacity for various small network topologies \cite{GeorgiadisTassiulas13,Wang10b,WangJournalCOPE}.

\begin{figure}
\centering
\subfigure[INC using only 3 operations\label{fig:BPEC_eg}]{
  \includegraphics[width=3.5cm]{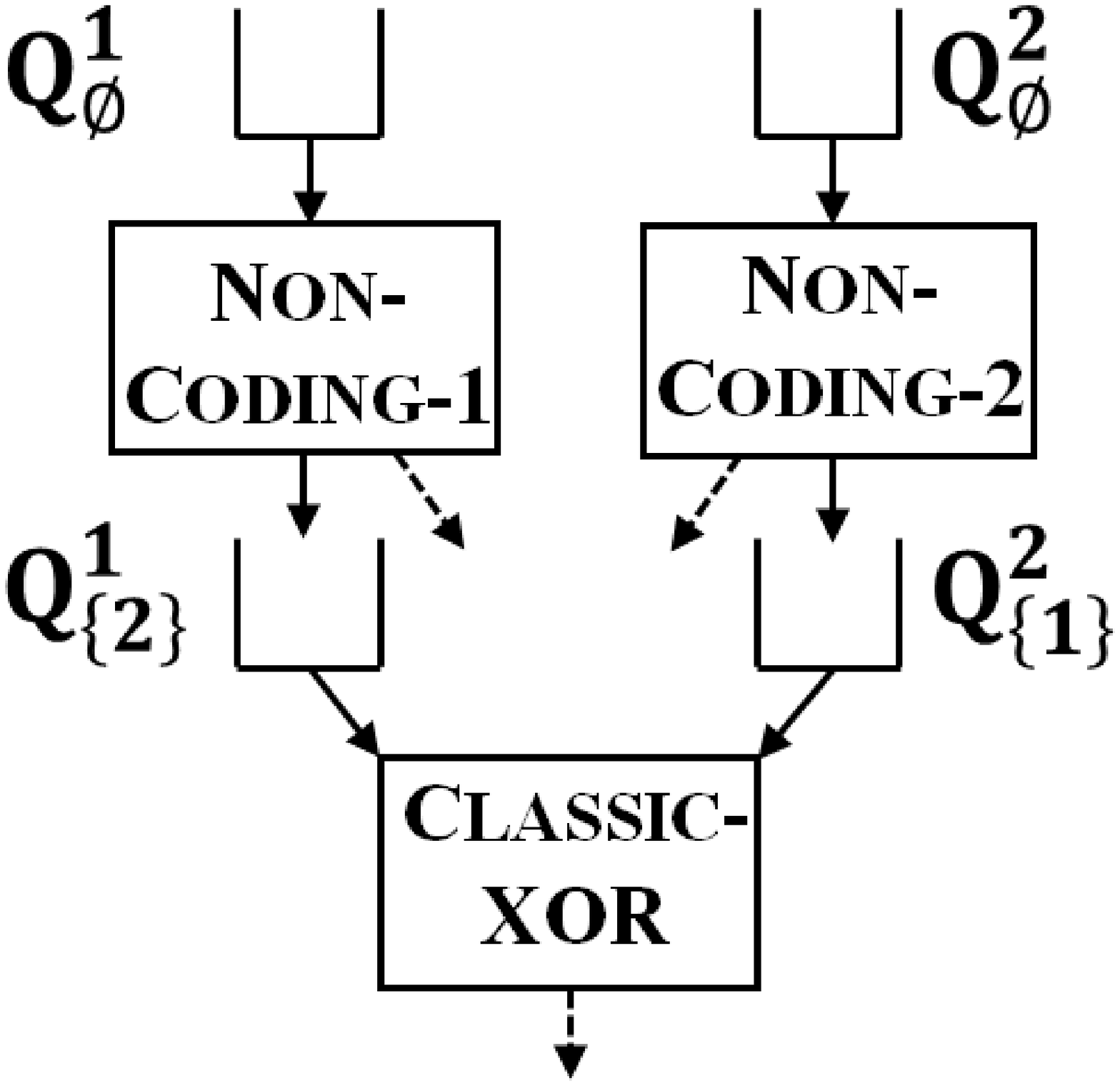}}
\subfigure[INC using only 5 operations\label{fig:5-type}]{
  \includegraphics[width=4.5cm]{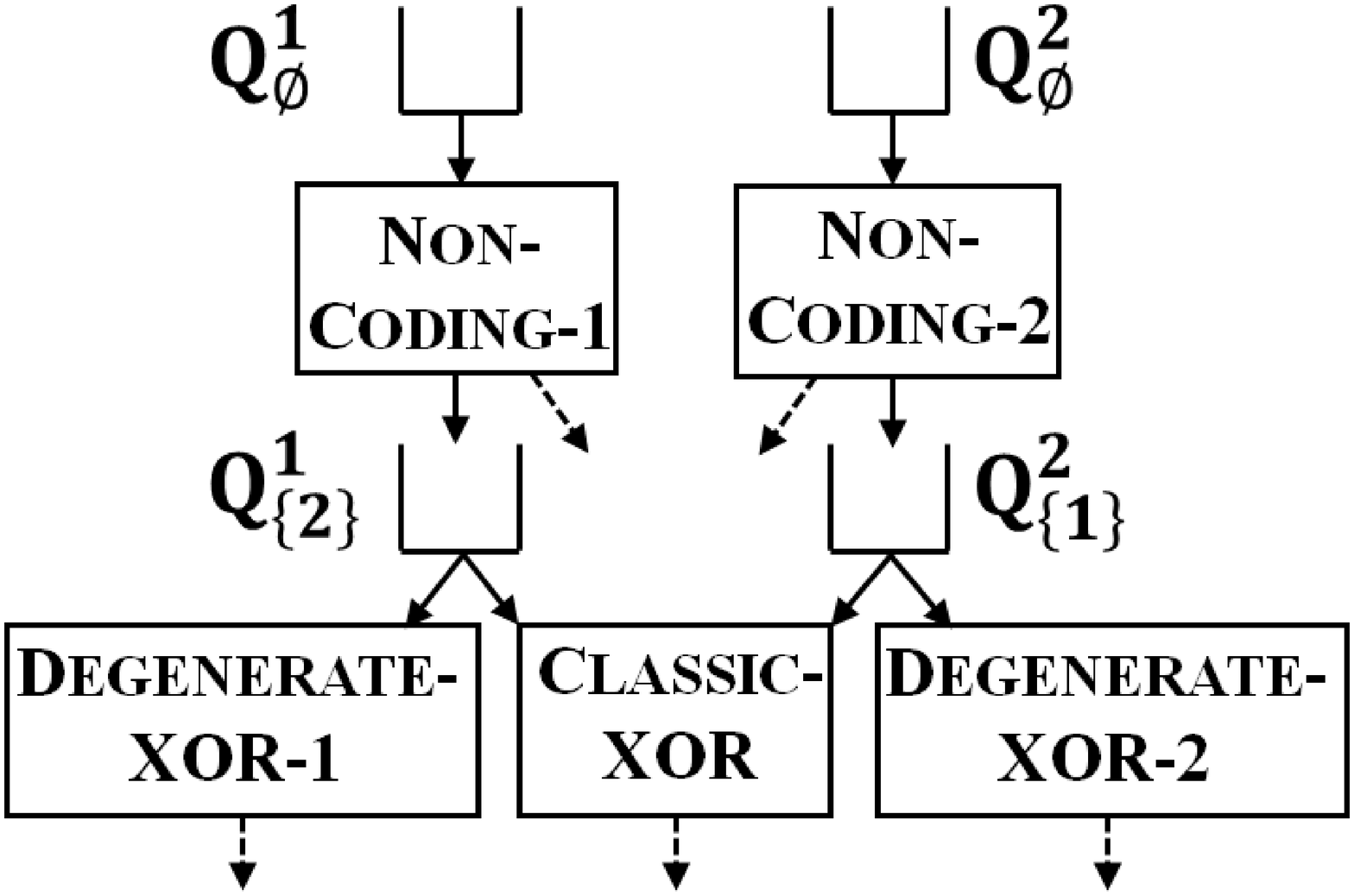}}  \caption{The virtual networks of two INC schemes.}
\vspace{-.5cm}
 \end{figure}
However, unlike the case of intra-session NC, there is no direct analogy from INC to the commodity flow. As a result, it is much more challenging to derive BP-based scheduling for INC traffic. We use the following example to illustrate this point. Consider a single source $s$ and two destinations $d_1$ and $d_2$. Source $s$ would like to send to $d_1$ the $X_i$ packets, $i=1,2,\cdots$; and send to $d_2$ the  $Y_j$ packets, $j=1,2,\cdots$. The simplest INC scheme consists of three operations. OP1: Send uncodedly those $X_i$ that have not been heard by any of $\{d_1,d_2\}$. OP2: Send uncodedly those $Y_j$ that have not been heard by any of $\{d_1,d_2\}$.  OP3: Send a linear sum $[X_i+Y_j]$ where $X_i$ has been overheard by $d_2$ but not by $d_1$ and $Y_j$ has been overheard by $d_1$ but not by $d_2$. For future reference, we denote OP1 to OP3 by  {\sc Non-Coding-1},  {\sc Non-Coding-2}, and {\sc Classic-XOR}, respectively.

OP1 to OP3 can also be represented by the virtual network (vr-network) in Fig.~\ref{fig:BPEC_eg}. Namely, any newly arrived $X_i$ and $Y_j$ virtual packets\footnote{We denote the packets (jobs) inside the vr-network by ``virtual packets.''} (vr-packets) that have not been heard by any of $\{d_1,d_2\}$ are stored in queues $Q^1_\emptyset$ and  $Q^2_\emptyset$, respectively. The superscript $k\in\{1,2\}$ indicates that the queue is for the session-$k$ packets. The subscript $\emptyset$ indicates that those packets have not been heard by any of $\{d_1,d_2\}$.  {\sc Non-Coding-1} then takes one $X_i$ vr-packet from $Q^1_\emptyset$ and send it uncodedly. If such $X_i$ is heard by $d_1$, then the vr-packet leaves the vr-network, which is described by the dotted arrow emanating from the {\sc Non-Coding-1} block. If $X_i$ is overheard by $d_2$ but not $d_1$, then we place it in queue $Q^1_{\{2\}}$, the queue for the overheard session-1 packets. {\sc Non-Coding-2} in Fig.~\ref{fig:BPEC_eg} can be interpreted symmetrically.  {\sc Classic-XOR} operation takes an $X_i$ from $Q^{1}_{\{2\}}$ and a $Y_j$ from $Q^2_{\{1\}}$ and sends $[X_i+Y_j]$. If $d_1$ receives $[X_i+Y_j]$, then $X_i$ is removed from $Q^{1}_{\{2\}}$ and leaves the vr-network. If $d_2$ receives $[X_i+Y_j]$, then $Y_j$ is removed from $Q^{2}_{\{1\}}$ and leaves the vr-network.

It is known \cite{Paschos12} that with dynamic packet arrivals, any INC scheme that (i) uses only these three operations and (ii) attains bounded decoding delay with rates $(R_1,R_2)$ can be converted to a scheduling solution that stabilizes the vr-network with rates $(R_1,R_2)$, and vice versa. {\em The INC design problem is thus converted to a vr-network scheduling problem.} To distinguish the above INC design for dynamical arrivals (the concept of stability regions) from the INC design assuming infinite backlog and decoding delay (the concept of the Shannon capacity), we term the former {\em the dynamic INC} design problem and the latter {\em the block-code INC} design problem.

\begin{figure}
\centering
  \includegraphics[width=6cm]{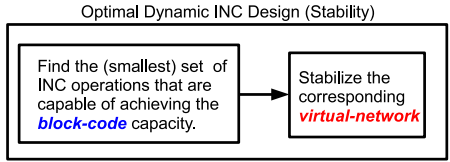}\\
  \caption{The two components of optimal dynamic INC design.}\label{fig:break-down}
  \vspace{-.3cm}
\end{figure}
The above vr-network representation also allows us to divide the optimal dynamic INC design problem into solving the following two major challenges separately. {\bf Challenge~1}: The example in Fig.~\ref{fig:BPEC_eg} focuses on dynamic INC schemes using only 3 possible operations. Obviously, the more INC operations one can choose from, the larger the degree of design freedom, and the higher the achievable throughput. {\em The goal is thus to find a (small) finite set of INC operations that can provably maximize the ``block-code'' achievable throughput}.  {\bf Challenge~2}: Suppose that we have found a set of INC operations that is capable of achieving the block-code capacity. However, it does not mean that such a set of INC operations will automatically lead to an optimal dynamic INC design since we still need to consider the delay/stability requirements. Specifically, once the best set of INC operations is decided, we can derive the corresponding vr-network as discussed in the previous paragraphs. {\em The goal then becomes to devise a stabilizing scheduling policy for the vr-network, which leads to an equivalent representation of the optimal dynamic INC solution.} See Fig.~\ref{fig:break-down} for the illustration of these two tasks.

Both tasks turn out to be highly non-trivial and optimal dynamic INC solution \cite{GeorgiadisTassiulas13,Paschos12,AthanasiadouGeorgiadis13} has been designed only for the scenario of fixed channel quality. Specifically, \cite{GeorgiadisTassiulas09} answers Challenge~1 and shows that for fixed channel quality,  the 3 INC operations in Fig.~\ref{fig:BPEC_eg} plus 2 additional {\sc Degenerate-XOR} operations, see Fig.~\ref{fig:5-type} and Section~\ref{subsubsec:degenerate-XOR}, can achieve the block-code INC capacity. One difficulty of resolving Challenge~2 is that an INC operation may involve multiple queues simultaneously, e.g., {\sc Classic-XOR} can only be scheduled when {\em both} $Q^1_{\{2\}}$ and $Q^2_{\{1\}}$ are non-empty. This is in sharp contrast with the traditional BP solutions\cite{ShizhenInfocom12,ShizhenInfocom14} in which each queue can act independently.\footnote{A critical assumption in [Section II C.1 \cite{tassiulas1992stability}] is that if two queues $Q_1$ and $Q_2$ can be {\em activated} at the same time, then we can also choose to activate only one of the queues if desired. This is not the case in the vr-network. E.g., {\sc Classic-XOR} activates both $Q^1_{\{2\}}$ and $Q^2_{\{1\}}$ but no coding operation in Fig.~\ref{fig:BPEC_eg} activates only one of $Q^1_{\{2\}}$ and $Q^2_{\{1\}}$.} For the vr-network in Fig.~\ref{fig:5-type},
\cite{GeorgiadisTassiulas13} circumvents this problem by designing a fixed priority rule that gives strict precedence to the {\sc Classic-XOR} operation. Alternatively, \cite{Paschos12} derives a BP scheduling scheme by noticing that the vr-network in Fig.~\ref{fig:5-type} can be decoupled into two vr-subnetworks (one for each data session) so that the queues in each of the  vr-subnetworks can be activated independently and the traditional BP results follow.

However, the channel quality varies over time for practical wireless downlink scenarios. Therefore, one should opportunistically choose the most favorable users as receivers, the so-called {\em opportunistic scheduling} technique. Recently \cite{WangHan14} shows that when allowing opportunistic coding+scheduling for time-varying channels, the 5 operations in Fig.~\ref{fig:5-type} no longer achieve the block-code capacity. The existing dynamic INC design in \cite{GeorgiadisTassiulas13,Paschos12} are thus strictly suboptimal for time-varying channels since they are based on a suboptimal set of INC operations (recall Fig.~\ref{fig:break-down}).

\ccwchange{This paper also considers {\em rate adaptation}.  When NC is not allowed, the existing practical schemes simply chooses} a reliable modulation-and-coding-scheme (MCS) (e.g. drop rate less than 0.1) with the highest transmission rate. However, when NC is allowed, it is not clear how to perform rate adaptation. The reason is that while using a high-rate MCS can directly increase the point-to-point throughput, using a low-rate MCS increases the chance of overhearing and thus maximizes the opportunity of performing {\sc Classic-XOR} that combines overheard packets to enhance throughput. How to balance the usage of high-rate and low-rate MCSs remained a critical and open problem in NC design.

This work proposes new optimal dynamic INC designs for 2-flow downlink traffic with time-varying packet erasure channels (PECs) and with rate adaptation. Our detailed contributions are summarized as follows.

{\em Contribution 1:} We introduce a new pair of INC operations such that (i) The underlying concept is distinctly different from the traditional wisdom of XORing two overheard packets; (ii) The overall scheme uses only the low-complexity binary XOR operation; and (iii) We prove that the new set of INC operations is capable of achieving the block-code-based Shannon capacity for the setting of time-varying PECs.

{\em Contribution 2:} The new INC operations lead to a new vr-network that is different from Fig.~\ref{fig:5-type} and the existing ``{\em vr-network decoupling} + {\em BP}'' approach in \cite{Paschos12} no longer holds. To answer Challenge~2, we generalize the results of Stochastic Processing Networks (SPNs) \cite{jiang2009stable,huang2011utility} and apply it to the new vr-network. The end result is an opportunistic, dynamic INC solution
that is {\em queue-length-based} and can robustly achieve the optimal stability region of time-varying PECs. 

{\em Contribution 3:} The proposed solution is also generalized for rate-adaptation. In simulations, our scheme can opportunistically and optimally choose the MCS of each packet transmission while achieving the optimal stability region, i.e., equal to the Shannon capacity. This new result is the first capacity-achieving INC solution which considers jointly coding, scheduling, and rate adaptation \change{for 1-base-station-2-session-client scenario}. 

{\em Contribution 4:} A byproduct of our results is a scheduling scheme for SPNs with {\em random departure} instead of {\em deterministic departure}, which relaxes a major limitation of the existing SPN model. The results could thus further broaden the applications of SPN scheduling to other real-world scenarios.

{\bf Organization of this work:} Section~\ref{sec:existing} defines the optimal stability region when allowing arbitrary NC operations. Sections~\ref{sec:new-coding} first explains the sub-optimality of existing INC operations and then introduce two new XOR-based operations that are capable of achieving the optimal Shannon capacity. The corresponding vr-network is also described in Section~\ref{sec:new-coding}. Section~\ref{sec:new_scheduling}  proposes a new scheduling scheme for the corresponding vr-network. Section~\ref{sec:combined} combines the new vr-network and the scheduling scheme and prove that the combined solution achieves the optimal stability region of any possible INC schemes.  In Sections~\ref{sec:existing} to \ref{sec:combined}, we focus exclusively on time-varying channels. In Section~\ref{subsec:extention_ACM}, we further generalize the proposed solution for rate adaptation and show numerically that it again achieves the optimal stability region.

{\bf Related Results:} The most related works are \cite{GeorgiadisTassiulas13,AthanasiadouGeorgiadis13,GeorgiadisTassiulas09,Paschos12}, which provide either a policy-based or a BP-based scheduling scheme for downlink networks. While they all achieve the 2-flow capacity of fixed channel quality, they are strictly suboptimal for time-varying PECs and for rate-adaptation. Other works \cite{PaschosFragiadakis13,KuoWangIT13} study the benefits of external side information with fixed channel quality and no rate-adaptation.

\section{Problem Formulation and Existing Results\label{sec:existing}}

\def\cq{{\mathsf{cq}}}
\def\bigcq{{\mathsf{CQ}}}
\def\stdef{\stackrel{\Delta}{=}}
\def\vp{{\vec{p}}}
\def\rcpt{{\mathsf{rcpt}}}

\subsection{Problem Formulation --- The Broadcast Erasure Channel\label{subsec:setting}}
\begin{figure}
\centering
  \includegraphics[width=5cm]{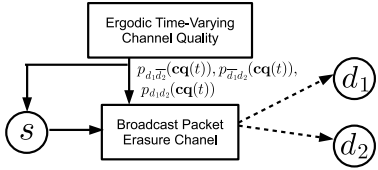}\\
  \caption{The time-varying broadcast packet erasure channel.}\label{fig:VCS}
\end{figure}

We model the 1-base-station/2-client downlink traffic as a broadcast packet erasure channel (PEC). See Fig.~\ref{fig:VCS} for illustration. The base station is sometimes called the source $s$. Consider the following slotted transmission system.

{\em Dynamic Arrival:} We assume that each incoming session-$i$ packet takes a value from a finite field $\GF(\varrho)$. In the beginning of every time slot $t$, there are $A_1(t)$ session-1 packets and $A_2(t)$ session-2 packets arriving at the source $s$. We assume that $A_1(t)$ and $A_2(t)$ are i.i.d.\ integer-valued random variables with mean $(\EE\{A_1(t)\},\EE\{A_2(t)\})=(R_1,R_2)$ and bounded support. Recall that $X_i$ and $Y_j$, $i,j\in\NN$, denote the session-1 and session-2 packets, respectively.

{\em Time-Varying Channel:}
We model the time-varying channel quality by a random process $\cq(t)$, which decides the reception probability of the {\em broadcast PEC.}  In all our proofs, we assume $\cq(t)$ is i.i.d.\ As will be seen, our scheme can be directly applied to Markovian $\cq(t)$ as well. Simulation shows that it also achieves the optimal stability region for Markovian $\cq(t)$ \cite{Wang10b} (albeit without any analytical proof). Due to space limits, the simulation results for Markovian $\cq(t)$ are omitted.

\par Let $\bigcq$ denote the support of $\cq(t)$ and we assume $|\bigcq|$ is finite. For any $c\in\bigcq$, we use $f_c$ to denote the steady state frequency of $\cq(t)=c$. We assume $f_c>0$ for all $c\in\bigcq$. 

{\em Broadcast Packet Erasure Channel:}
For each time slot $t$, source $s$ can transmit one packet, $\pkt(t)\in \GF(\varrho)$, which will be received by a random subset of destinations $\{d_1,d_2\}$, and let $\rcvpkt{i}(t)\in\{\pkt(t),*\}$ denote the received packet at destination $d_i$ in time $t$. That is, $\rcvpkt{i}(t)=\pkt(t)$ means that the packet is received successfully and $\rcvpkt{i}(t)=*$, the erasure symbol, means that the received packet is corrupted and discarded completely. Specifically, there are 4 possible {\em reception status} $\{\overline{d_1d_2},d_1\overline{d_2},\overline{d_1}d_2,d_1d_2\}$, e.g., the reception status $\rcpt=d_1\overline{d_2}$ means that the packet is received by $d_1$ but not $d_2$. The reception status probabilities can be described by a vector $\vec{p}\stdef(p_{\overline{d_1d_2}}, p_{d_1\overline{d_2}},p_{\overline{d_1}d_2},p_{d_1d_2})$. For example, $\vp=(0,0.5,0.5,0)$ means that every time we transmit a packet, with 0.5 probability it will be received by $d_1$ only and with 0.5 probability it will be received by $d_2$ only. In contrast, if we have $\vp=(0,0,0,1)$, then it means that the packet is always received by $d_1$ and $d_2$ simultaneously. Since our model allows arbitrary joint probability vector $\vec{p}$, it captures the scenarios in which the erasure events of $d_1$ and $d_2$ are dependent, e.g., when the erasures at $d_1$ and $d_2$ are caused by a common (random)  interference source.

{\em Opportunistic INC:}
Since the reception probability is decided by the channel quality, we write $\vp(\cq(t))$ as a function of $\cq(t)$ at time $t$. In the beginning of time $t$, we assume that $s$ is aware of the channel quality $\cq(t)$ (and thus knows $\vp(\cq(t))$) so that $s$ can opportunistically decide how to encode the packet for time $t$. See Fig.~\ref{fig:VCS}. This is motivated by Cognitive Radio, for which $s$ can sense the channel first before transmission.

{\em ACKnowledgement:} In the end of time $t$, $d_1$ and $d_2$ will report back to $s$ whether they have received the transmitted packet or not (ACK/NACK). A useful notation regarding the ACK feedback is as follows.
We use a 2-dimensional {\em channel status vector} $\Z(t)$ to represent the channel reception status:
\begin{align}
\Z(t) = (Z_{d_1}(t),Z_{d_2}(t)) \in \{*,1\}^2 \nonumber
\end{align}
where ``$*$" and ``1" represent erasure and successful reception, respectively. For example, when $s$ transmits a packet $\pkt(t)\in\GF(\varrho)$ in time $t$, the destination $d_1$ receives $\rcvpkt{1}(t)=\pkt(t)$ if $Z_{d_1}(t)=1$, and receives $\rcvpkt{1}(t)=*$ if $Z_{d_1}(t)=*$.

{\em Buffers:}
There are two buffers at $s$ which stores the incoming session-1 packets and session-2 packets, respectively. Let $\buf_i(t)$ denote the collection of all session-$i$ packets currently stored in buffer $i$ in the beginning of time $t$.

{\em Encoding, Buffer Management, and Decoding:} For simplicity, we use $[\cdot]_1^t$ to denote the collection from time $1$ to $t$. For example, $[A_1,\Z]_1^t\eqdef \{A_1(\tau), \Z(\tau):\forall \tau\in\{1,2,...,t\}\}$. At each time $t$, an opportunistic INC solution is defined by an {\em encoding function} and two {\em buffer pruning functions:} 

{\bf Encoding:} In the {\em beginning} of time $t$, the coded packet $W(t)$ sent by $s$ is expressed by
\begin{align}
\pkt(t) = f_{\textsf{ENC},t}(\buf_1(t),\buf_2(t),[\Z]_1^{t-1}). \label{eq:pkt_encoding}
\end{align}
That is, the coded packet is generated by the packets that are still in the two buffers and by the past reception status feedback from time 1 to $t-1$.

{\bf Buffer Management:} In the {\em end} of time $t$, we prune the buffers by the following equation: For $i=1,2$,
\begin{align}
\buf_i(t+1) &= \buf_i(t) \backslash f_{\textsf{PRUNE},i,t}([A_1, A_2,\Z]_1^t) \nonumber \\
&\hspace{-1cm}\cup \{\text{New session-$i$ packets arrived in time $t$}\}. \label{eq:buffer_pruning}
\end{align}
That is, the buffer pruning function $f_{\textsf{PRUNE},i,t}([A_1,A_2,\Z]_1^t)$ will decide which packets to remove from $\buf_i(t)$ based on the arrival and the packet delivery patterns from time 1 to $t$, while new packets will also be stored in the buffer. $\buf_i(t+1)$ will later be used for encoding in time $t+1$.

\par The encoding and the buffer pruning functions need to satisfy the following {\bf Decodability Condition:} For every time $t$, there exist two decoding functions such that
\begin{align}
(X_1,...,X_{\sum_{\tau=1}^{t}A_1(\tau)}) &= f_{\textsf{DEC},1,t}([\rcvpkt{1}]_1^{t},\buf_1(t+1))\nonumber \\
(Y_1,...,Y_{\sum_{\tau=1}^{t}A_2(\tau)}) &= f_{\textsf{DEC},2,t}([\rcvpkt{2}]_1^{t},\buf_2(t+1)). \label{eq:pkt_dec2}
\end{align}

The intuition of the above decodability requirement \eqref{eq:pkt_dec2} is as follows. If the pruning function \eqref{eq:buffer_pruning} is very aggressive, then the buffer size at source $s$ is small but there is some risk that some desired messages $X_i$ may be ``removed from $\buf_1(t)$ prematurely.'' That is, once $X_i$ is removed, it can no longer be decoded at $d_1$ even if we send all the content in the remaining $\buf_1(t+1)$ directly to $d_1$. To avoid this undesired consequence, \eqref{eq:pkt_dec2} imposes that the pruning function has to be conservative in the sense that if in the end of time $t$ we let $d_i$ directly access $\buf_i(t+1)$ at the source $s$, then together with what $d_i$ has already received $[\rcvpkt{i}]_1^t$, $d_i$ should be able to fully recover all the session-$i$ packets up to time $t$.

\begin{definition}
A queue length $q(t)$ is {\em mean-rate stable} \cite{MJNeely12} (sometimes known as sublinearly stable)
if 
\begin{align}
\lim_{t\to\infty}\frac{\EE\{|q(t)|\}}{t}=0. \label{eq:stability-region-def}
\end{align}
\end{definition}

\begin{definition}\label{def:mean-rate} An arrival rate vector $(R_1, R_2)$ is {\em mean-rate stable} if there exists an NC scheme described by $f_{\textsf{ENC},t}$, $f_{\textsf{PRUNE},1,t}$ and $f_{\textsf{PRUNE},2,t}$ (which have to satisfy \eqref{eq:pkt_dec2}) such that the sizes of $\buf_1(t)$ and $\buf_2(t)$ are mean-rate stable. The NC stability region is the collection of all mean-rate stable vectors $(R_1,R_2)$.
\end{definition}
\par The above definition of mean-rate stability is a strict generalization of the traditional stability definition of uncoded transmissions. For example, suppose we decide to {\em not} using NC. Then we simply set the encoder $f_{\textsf{ENC},t}$ to always return either ``$X_i$ for some $i$'' or ``$Y_j$ for some $j$''. And we prune $X_i$ from $\buf_1(t)$ (resp.\  $Y_j$ from $\buf_2(t)$) if $X_i$ (resp.\ $Y_j$) was delivered successfully to $d_1$ (resp.\ $d_2$). The decodability condition \eqref{eq:pkt_dec2} holds naturally for the above $f_{\textsf{ENC},t}$, $f_{\textsf{PRUNE},1,t}$ and $f_{\textsf{PRUNE},2,t}$ and the buffers are basically the packet queues in the traditional non-coding schemes. Our new stability definition is thus equivalent to the traditional one once we restrict to non-coding solutions only.

On the other hand, when NC is allowed, the situation changes significantly. For example, an arbitrary NC scheme may send a coded packet that is a linear sum of three packets, say $[X_3+X_5+Y_4]$. Suppose the linear sum is received by $d_1$. The NC scheme then needs to carefully decide whether to remove $X_3$ or $X_5$ or both from $\buf_1(t)$ and/or whether to remove $Y_4$ from $\buf_2(t)$. 
This is the reason why we specify an NC scheme not only by the encoder $f_{\textsf{ENC},t}$ but also by the buffer management policy $f_{\textsf{PRUNE},1,t}$ and $f_{\textsf{PRUNE},2,t}$. Since our stability definition allows for arbitrary $f_{\textsf{ENC},t}$, $f_{\textsf{PRUNE},k,t}$ and $f_{\textsf{PRUNE},2,t}$, it thus represents the largest possible stability region that can be achieved by any NC solutions.


\subsection{Shannon Capacity Region}
Reference \cite{WangHan14} focuses on the above setting but considers the infinite backlog block-code design. 
We summarize the Shannon capacity result in \cite{WangHan14} as follows.

\begin{proposition}\label{prop:recite}[Propositions~1 and 3, \cite{WangHan14}] For the block-code setting with infinite backlog, the closure of all achievable rate vectors $(R_1,R_2)$ can be characterized by $|\bigcq|+12$ linear inequalities that involve $18\cdot|\bigcq|+7$ non-negative auxiliary variables. As a result, the Shannon capacity region $(R_1,R_2)$ can be explicitly computed by solving the corresponding LP problem. Detailed description of the LP problem can be found in \cite{WangHan14,KuoWang14:techrepstdsub}.
\end{proposition}

Since the block-code setting is less stringent than the dynamic arrival setting in this work, the above Shannon capacity region serves as an outer bound for the mean-rate stability region in Definition~\ref{def:mean-rate}. Our goal is to design a dynamic INC scheme, of which the stability region matches the Shannon capacity region.

\section{The Proposed New INC Operations} \label{sec:new-coding}
In this section, we aim to solve Challenge 1 in Section~\ref{sec:introduction}. We first discuss the limitations of the existing works on the INC block code design. We then describe a new set of binary INC operations that is capable of achieving the block code capacity. As discussed in Section~\ref{sec:introduction} and Fig.~\ref{fig:break-down}, knowing the best set of INC operations alone is not enough to achieve the largest stability region. Our new virtual network scheduler design will be presented separately in Section~\ref{sec:new_scheduling}.

\subsection{The 5 INC operations are no longer optimal\label{subsubsec:degenerate-XOR}}

In Section~\ref{sec:introduction}, we have detailed 3 INC operations: {\sc Non-Coding-1}, {\sc Non-Coding-2}, and {\sc Classic-XOR}. Two additional INC operations are introduced in \cite{GeorgiadisTassiulas09}: {\sc Degenerate-XOR-1} and {\sc Degenerate-XOR-2} as illustrated in Fig.~\ref{fig:5-type}. Specifically, {\sc Degenerate-XOR-1} is designed to handle the degenerate case in which $Q^1_{\{2\}}$ is non empty but  $Q^2_{\{1\}}=\emptyset$. Namely, there is at least one $X_i$ packet overheard by $d_2$ but there is no $Y_j$ packet overheard by $d_1$. Not having such $Y_j$ implies that one cannot send $[X_i+Y_j]$ (the {\sc Classic-XOR} operation). An alternative is thus to send the overheard $X_i$ uncodedly (as if sending $[X_i+0]$). We term this operation {\sc Degenerate-XOR-1}. One can see from Fig.~\ref{fig:5-type} that {\sc Degenerate-XOR-1} takes a vr-packet from $Q^1_{\{2\}}$ as input. If $d_1$ receives it, the vr-packet will leave the vr-network. {\sc Degenerate-XOR-2} is the symmetric version of {\sc Degenerate-XOR-1}.

We use the following example to illustrate the sub-optimality of the above 5 operations. Suppose $s$ has an $X$ packet for $d_1$ and a $Y$ packet for $d_2$ and consider a duration of 2 time slots. Also suppose that $s$ knows beforehand that the time-varying channel will have (i) $\vp=(0,0.5,0.5,0)$ for slot~1; and (ii) $\vp=(0,0,0,1)$ for slot~2. The goal is to transmit as many packets in 2 time slots as possible.

{\em Solution~1: INC based on the 5 operations in Fig.~\ref{fig:5-type}.}  In the beginning of time 1, both $Q^1_{\{2\}}$ and $Q^2_{\{1\}}$ are empty. Therefore, we can only choose either {\sc Non-Coding-1} or {\sc Non-Coding-2}. Without loss of generality we choose {\sc Non-Coding-1} and send $X$ uncodedly. Since $\vp=(0,0.5,0.5,0)$ in slot~1, there are only two cases to consider. Case 1: $X$ is received only by $d_1$. In this case, we can send $Y$ in the second time slot, which is guaranteed to arrive at $d_2$ since $\vp=(0,0,0,1)$ in slot~2. The total sum rate is sending 2 packets ($X$ and $Y$) in 2 time slots.
Case 2: $X$ is received only by $d_2$. In this case, $Q^1_{\{2\}}$ contains one packet $X$, and $Q^2_\emptyset$ contains one packet $Y$, and all the other queues in Fig.~\ref{fig:5-type} are empty. We can thus choose either {\sc Non-Coding-2} or {\sc Degenerate-XOR-1} for slot~2. Regardless of which coding operation we choose, slot~2 will then deliver 1 packet to either $d_2$ or $d_1$, depending on the INC operation we choose. Since no packet is delivered in slot~1, the total sum rate is 1 packet in 2 time slots. Since both cases have probability 0.5, the expected sum rate is $2\cdot 0.5 + 1\cdot 0.5=1.5$ packets in 2 time slots.

{\em An optimal solution:} We can achieve strictly better throughput by introducing new INC operations. Specifically, in slot~1, we send the linear sum $[X+Y]$ {\em even though neither $X$ nor $Y$ has ever been transmitted}, a distinct departure from the existing 5-operation-based solutions.

Again consider two cases:  Case 1: $[X+Y]$ is received only by $d_1$. In this case, we let $s$ send $Y$ uncodedly in slot~2. Since $\vp=(0,0,0,1)$ in slot~2, $Y$ will be received by both $d_1$ and $d_2$. $d_2$ is happy since it has now received the desired $Y$ packet. $d_1$ can use $Y$ together with the $[X+Y]$ packet received in slot~1 to decode its desired $X$ packet. Therefore, we deliver 2 packets ($X$ and $Y$) in 2 time slots. Case 2: $[X+Y]$ is received only by $d_2$. In this case, we let $s$ send $X$ uncodedly in slot~2. By the symmetric arguments, we deliver 2 packets ($X$ and $Y$) in 2 time slots.
The sum-rate of the new solution is 2 packets in 2 slots, a 33\% improvement over the existing solution.

{\em Remark:} This example focuses on a 2-time-slot duration due to the simplicity of the analysis. It is worth noting that the throughput improvement persists even for infinitely many time slots. See the simulation results in Section~\ref{sec:simulation}.

\begin{figure}
\centering
  \includegraphics[width=7cm]{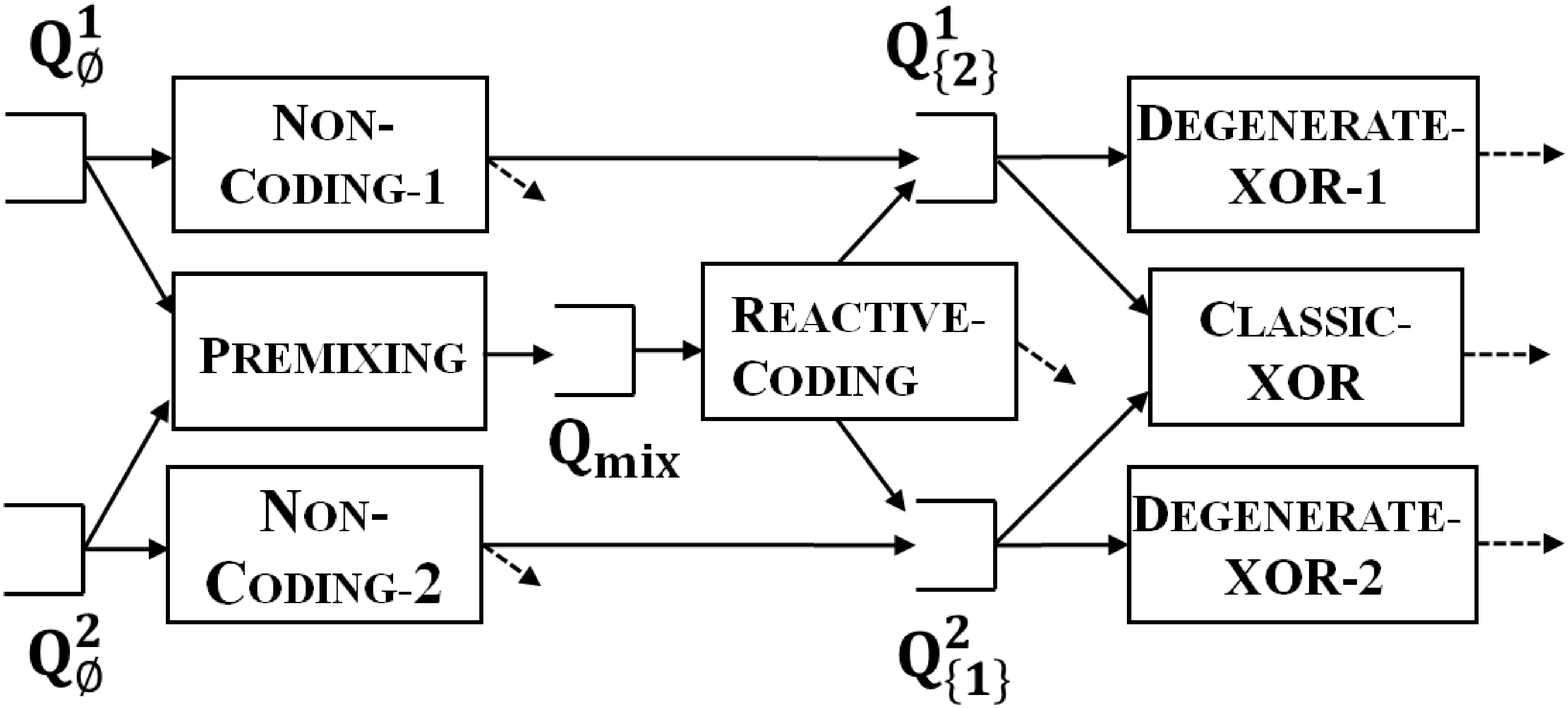}\\
  \caption{The virtual network of the proposed new INC solution.}\label{fig:7-type}
  \vspace{-.3cm}
\end{figure}

The above example shows that the set of 5 INC operations: {\sc Non-Coding-1}, {\sc Non-Coding-2}, {\sc Classic-XOR}, {\sc Degenerate-XOR-1}, and {\sc Degenerate-XOR-2} is not capable of achieving the Shannon capacity. To mitigate this inefficiency, we will enlarge the above set by introducing two more INC operations. We will first describe the corresponding encoder and then discuss the decoder and buffer management.

\subsection{Encoding Steps} \label{subsec:encoding}
We start from Fig.~\ref{fig:5-type}, the vr-network corresponding to the existing 5 INC operations. We then add 2 more operations, termed {\sc Premixing} and {\sc Reactive-Coding}, respectively, and 1 new virtual queue, termed $Q_\text{mix}$, and plot the vr-network of the new scheme in Fig.~\ref{fig:7-type}. From Fig.~\ref{fig:7-type}, we can clearly see that {\sc Premixing} involves both $Q^1_\emptyset$ and $Q^2_\emptyset$ as input and outputs to $Q_\text{mix}$. {\sc Reactive-Coding} involves $Q_\text{mix}$ as input and outputs to $Q^1_{\{2\}}$ or $Q^2_{\{1\}}$ or simply lets the vr-packet leave the vr-network (described by the dotted arrow).

\par In the following, we describe in detail how these two new INC operations 
work and how to integrate them with the other 5 operations. Our description contains 4 parts.

{\em Part I:}
The two operations, {\sc Non-Coding-1} and {\sc Non-Coding-2}, remain the same. That is, if we choose {\sc Non-Coding-1}, then $s$ chooses an uncoded session-1 packet $X_i$ from $Q^1_\emptyset$ and send it out. {\sc Non-Coding-2} is symmetric.

{\em Part II:} We now describe the new operation {\sc Premixing}. We can choose {\sc Premixing} only if both $Q^1_\emptyset$ and $Q^2_\emptyset$ are non-empty. Namely, there are $\{X_i\}$ packets and $\{Y_j\}$ packets that have not been heard by any of $d_1$ and $d_2$. Whenever we schedule {\sc Premixing}, we choose one $X_i$ from $Q^1_\emptyset$ and one $Y_j$ from $Q^2_\emptyset$ and send $[X_i+Y_j]$. If neither $d_1$ nor $d_2$ receives it, both $X_i$ and $Y_j$ remain in their original queues.

If at least one of $\{d_1,d_2\}$ receives it, we remove {\em both} $X_i$ and $Y_j$ from their queues and insert a tuple $(\rcpt; X_i,Y_j)$ into $Q_{\text{mix}}$. That is, unlike the other queues for which each entry is a single vr-packet, each entry of $Q_{\text{mix}}$ is a tuple.

The first coordinate of $(\rcpt; X_i,Y_j)$ is $\rcpt$, the reception status of $[X_i+Y_j]$. For example, if $[X_i+Y_j]$ was received by $d_2$ but not by $d_1$, then we set/record $\rcpt=\overline{d_1}d_2$; If $[X_i+Y_j]$ was received by both $d_1$ and $d_2$, then $\rcpt=d_1d_2$. The second and third coordinates store the participating packets $X_i$ and $Y_j$ separately. The reason why we do not store the linear sum directly is due to the new {\sc Reactive-Coding} operation.

{\em Part III:} We now describe the new operation {\sc Reactive-Coding}.
For any time $t$, we can choose {\sc Reactive-Coding} only if there is at least one tuple $(\rcpt; X_i,Y_j)$ in $Q_\text{mix}$. Choose one tuple from $Q_\text{mix}$ and denote it by $(\rcpt^*; X_i^*, Y_j^*)$. We now describe the encoding part of {\sc Reactive-Coding}.

If $\rcpt^*=d_1\overline{d_2}$, we send $Y_j^*$. If $\rcpt^*=\overline{d_1}d_2$ or ${d_1}d_2$, we send $X_i^*$. One can see that the coding operation depends on the reception status $\rcpt^*$ when $[X_i^*+Y_j^*]$ was first transmitted. This is why it is named {\sc Reactive-Coding}.

\begin{table}
\caption{A summary of the {\sc Reactive-Coding} operation\label{tab:reactive}}
\centering
 \includegraphics[width=8.5cm]{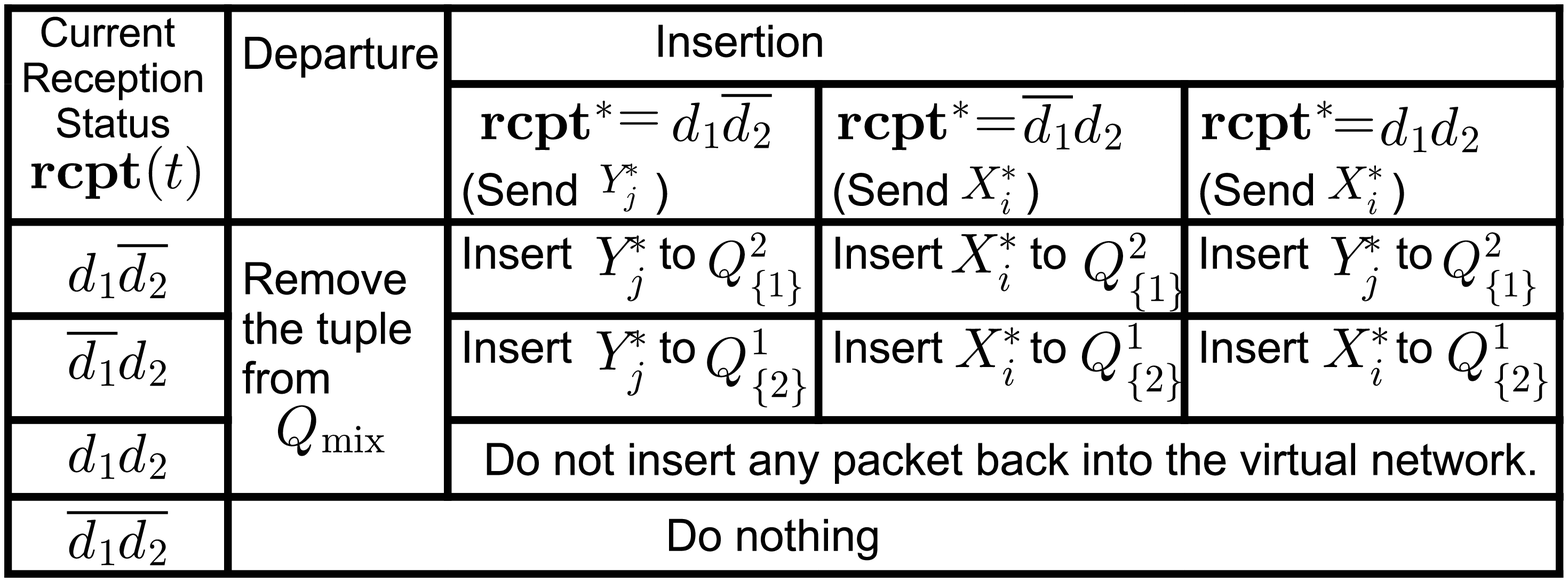}
 \vspace{-.0cm}
\end{table}

The movement of the vr-packets depends on the current reception status of time $t$, denoted by $\rcpt(t)$, and also on the old reception status $\rcpt^*$ when the sum $[X_i^*+Y_j^*]$ was originally transmitted. The detailed movement rules are described in Table~\ref{tab:reactive}. The way to interpret the table is as follows. When $\rcpt(t)=\overline{d_1d_2}$, i.e., neither $d_1$ nor $d_2$ receives the current transmission, then we do nothing, i.e., keep the tuple inside $Q_{\text{mix}}$. On the other hand, we remove the tuple from $Q_\text{mix}$ whenever $\rcpt(t)\in\{d_1\overline{d_2},\overline{d_1}d_2,d_1d_2\}$. If $\rcpt(t)=d_1d_2$, then we remove the tuple but do not insert any vr-packet back to the vr-network, see the second last row of Table~\ref{tab:reactive}. The tuple essentially leaves the vr-network in this case. If $\rcpt(t)=d_1\overline{d_2}$ and $\rcpt^*=d_1d_2$, then we remove the tuple from $Q_\text{mix}$ and insert $Y_j^*$ to $Q^2_{\{1\}}$. The rest of the combinations can be read from Table~\ref{tab:reactive} in the same way. One can verify that the optimal INC example introduced in Section~\ref{subsubsec:degenerate-XOR} is a direct application of the {\sc Premixing} and {\sc Reactive-Coding} operations.

Before proceeding, 
we briefly explain why the combination of {\sc Premixing} and {\sc Reactive-Coding} works.  To facilitate discussion, we call the time slot in which we use {\sc Premixing} to transmit $[X_i^*+Y_j^*]$ ``slot~1'' and the time slot in which we use {\sc Reactive-Coding} ``slot~2,'' even though the coding operations {\sc Premixing} and {\sc Reactive-Coding} may not be scheduled in two adjacent time slots. Using this notation, if  $\rcpt^*=d_1\overline{d_2}$ and $\rcpt(t)=d_1d_2$, then it means that $d_1$ receives $[X_i^*+Y_j^*]$ and $Y_j^*$ in slots~1 and 2, respectively and $d_2$ receives $Y_j^*$ in slot 2. In this case, $d_1$ can decode the desired $X_i^*$ and $d_2$ directly receives the desired $Y_j^*$. We now consider the perspective of the vr-network. Table~\ref{tab:reactive} shows that the tuple will be removed from $Q_\text{mix}$ and leave the vr-network. Therefore, no queue in the vr-network stores any of $X_i^*$ and $Y_j^*$. This correctly reflects the fact that both $X_i^*$ and $Y_j^*$ have been received by their intended destinations.

Another example is when $\rcpt^*=\overline{d_1}{d_2}$ and $\rcpt(t)=d_1\overline{d_2}$. In this case,  $d_2$ receives $[X_i^*+Y_j^*]$ in slot 1 and $d_1$ receives $X_i^*$ in slot~2.  From the vr-network's perspective, the movement rule (see Table~\ref{tab:reactive}) removes the tuple from $Q_\text{mix}$ and insert an $X_i^*$ packet to $Q^2_{\{1\}}$. Since a vr-packet is removed from a session-1 queue\footnote{$Q_\text{mix}$ is regarded as both a session-1 and a session-2 queue simultaneously.} $Q_\text{mix}$ and inserted to a session-2 queue $Q^2_{\{1\}}$,  the total number of vr-packets in the session-1 queue decreases by 1. This correctly reflects the fact that $d_1$ has received 1 desired packet $X_i^*$ in slot~2.

An astute reader may wonder why in this example we can put $X_i^*$, a session-1 packet, into a session-2 queue $Q^2_{\{1\}}$. The reason is that whenever $d_2$ receives $X_i^*$ in the future, it can recover its desired $Y_j^*$ by subtracting $X_i^*$ from the linear sum $[X_i^*+Y_j^*]$ it received in slot 1 (recall that $\rcpt^*=d_1\overline{d_2}$.) Therefore, $X_i^*$ is now information-equivalent to $Y_j^*$, a session-2 packet. Moreover, $d_1$ has received $X_i^*$. Therefore, in terms of the information it carries, $X_i^*$ is no different than a session-2 packet that has been overheard   by $d_1$. As a result, it is fit to put $X_i^*$ in $Q^2_{\{1\}}$.

{\em Part IV:} We now describe some slight modification to {\sc Classic-XOR}, {\sc Degenerate-XOR-1}, and {\sc Degenerate-XOR-2}. A unique feature of the new scheme is that some packets in $Q^2_{\{1\}}$ may be an $X_i^*$ packet that is inserted by {\sc Reactive-Coding} when $\rcpt^*=\overline{d_1}{d_2}$ and $\rcpt(t)=d_1\overline{d_2}$. (Also some $Q^1_{\{2\}}$ packets may be $Y_j^*$.) However, in our previous discussion, we have shown such an $X_i^*$ in $Q^2_{\{1\}}$ is information-equivalent to a $Y_j^*$ packet overheard by $d_1$. Therefore, in the {\sc Classic-XOR} operation, we should not insist on sending $[X_i+Y_j]$ but can also send $[P_1+P_2]$ as long as $P_1$ is from $Q^1_{\{2\}}$ and $P_2$ is from $Q^2_{\{1\}}$. The same relaxation must be applied to {\sc Degenerate-XOR-1} and {\sc Degenerate-XOR-2} operations. Other than this slight relaxation, the three operations work in the same way as previously described in Sections~\ref{sec:introduction} and \ref{subsubsec:degenerate-XOR}.


We conclude this section by listing in Table~\ref{tab:transition} the transition probabilities of half of the edges of the vr-network in Fig.~\ref{fig:7-type}. E.g., when we schedule {\sc Premixing}, we remove a packet from $Q^1_\emptyset$ if at least one of $\{d_1,d_2\}$ receives it. The transition probability along the $Q^1_\emptyset\rightarrow${\sc Premixing} edge is thus $p_{d_1\vee d_2}\stdef p_{d_1\overline{d_2}}+p_{\overline{d_1}d_2}+p_{d_1d_2}$. All the other transition probabilities in Table~\ref{tab:transition} can be derived similarly. The transition probability of the rest of the edges can be derived by symmetry.

\begin{table}
\caption{A summary of the transition probability of the virtual network in Fig.~\ref{fig:7-type}, where $p_{d_1\vee d_2}\stdef p_{d_1\overline{d_2}}+p_{\overline{d_1}d_2}+p_{d_1d_2}$; $p_{d_1}\stdef p_{d_1\overline{d_2}}+p_{d_1{d_2}}$; {\sc NC1} stands for {\sc Non-Coding-1}; {\sc CX} stands for {\sc Classic-XOR}; {\sc DX1} stands for {\sc Degenerate-XOR-1}; {\sc PM} stands for {\sc Premixing}; {\sc RC} stands for {\sc Reactive-Coding}. 
\label{tab:transition}}
\centering
\begin{tabular}{cl|cl}
\hline
\hline
Edge & Trans.\ Prob. & Edge & Trans.\ Prob.\\
\hline
$Q^1_\emptyset\rightarrow${\sc NC1} & $p_{d_1\vee d_2}$ & $Q^1_\emptyset\rightarrow${\sc PM} & $p_{d_1\vee d_2}$ \\
{\sc NC1}$\rightarrow Q^1_{\{2\}}$ & $p_{\overline{d_1}d_2}$ & {\sc PM}$\rightarrow Q_\text{mix}$& $p_{d_1\vee d_2}$ \\
$Q^1_{\{2\}}\rightarrow${\sc DX1} & $p_{d_1}$ & $Q_\text{mix}\rightarrow${\sc RC} & $p_{d_1\vee d_2}$ \\
$Q^1_{\{2\}}\rightarrow${\sc CX} & $p_{d_1}$ & {\sc RC}$\rightarrow Q^1_{\{2\}}$& $p_{\overline{d_1}d_2}$\\
\hline
\hline
 \end{tabular}
\end{table}

\subsection{Decoding and Buffer Management at Receivers} \label{subsec:buf_mang_rx}

\par The vr-network is a conceptual tool used by $s$ to decide what to transmit in each time slot. As a result, for encoding, $s$ only needs to store in its memory all the packets that are currently in the vr-network. This implies that as long as the queues in the vr-network are stable, the actual memory usage (buffer size) at the source is also stable. However, one also needs to ensure that the memory usage for receivers is stable as well. In this subsection we discuss the decoding operations and the memory usage at the receivers.

\par 
A very commonly used assumption in the Shannon-capacity literature is to assume that the receivers store all the overheard packets so that they can use them to decode any XORed packets sent from the source. No packets will ever be removed from the buffer under such a policy. Obviously, such an infinite-buffer scheme is highly impractical.

\par When there is only 1 session in the network, Gaussian elimination (GE) is often used for buffer management. However, generalizing GE for the multi-session non-generation-based schemes can be very complicated 
\cite{KoutWangHuShroff11}.

\par In the existing multi-session INC works \cite{KattiRahulHuKatabiMedardCrowcroft06,GeorgiadisTassiulas13,AthanasiadouGeorgiadis13,Paschos12}, a commonly used buffer management scheme is the following. For any time $t$, define $i^*$ (resp. $j^*$) as the smallest $i$ (resp. $j$) such that $d_1$ (resp. $d_2$) has not decoded $X_i$ (resp. $Y_j$) in the end of time $t$. Then each receiver can simply remove any $X_i$ and $Y_j$ from its buffer for those $i<i^*$ and $j<j^*$. The reason is that since those $X_i$ and $Y_j$ have already been known by their intended receivers, they will not participate in any future transmission, and thus can be removed from the buffer. 

\par On the other hand, under such a buffer management scheme, the receivers may use significantly more memory than that of the source. 
The reason is as follows. Suppose $d_1$ has decoded $X_1$, $X_3$, $X_4,\cdots,X_8$, and $X_{10}$ and suppose $d_2$ has decoded $Y_1$ to $Y_4$ and $Y_6$ to $Y_{10}$. In this case $i^*=2$ and $j^*=5$. The aforementioned scheme will keep all $X_2$ to $X_{10}$ in the buffer of $d_2$ and all $Y_5$ to $Y_{10}$ in the buffer of $d_1$ even though the source is interested in only sending 3 more packets $X_2$, $X_9$, and $Y_5$. The above buffer management scheme is too conservative since it does not trace the actual overhearing status of each packet and only use  $i^*$ and $j^*$ to decide whether to prune the packets in the buffers of the receivers.


\par In contrast, 
our vr-network scheme admits the following efficient decoding operations and buffer management. In the following, we describe the decoding and buffer management at $d_1$. The operations at $d_2$ can be done symmetrically. Our description consists of two parts. We first describe how to perform decoding at $d_1$ and which packets need to be stored in $d_1$'s buffer, {\em while assuming that any packets that have ever been stored in the buffer will never be expunged.} In the second part, we describe how to prune the memory usage without affecting the decoding operations.

{\bf Upon $d_1$ receiving a packet:} Case 1: If the received packet is generated by {\sc Non-Coding-1}, then such a packet must be $X_i$ for some $i$. We thus pass such an $X_i$ to the upper layer; Case 2: If the received packet is generated by {\sc Non-Coding-2}, then such a packet must be $Y_j$ for some $j$. We store $Y_j$ in the buffer of $d_1$; Case 3: If the received packet is generated by {\sc Premixing}, then such a packet must be $[X_i+Y_j]$. We store the linear sum $[X_i+Y_j]$ in the buffer. Case 4: If the received packet is generated by {\sc Reactive Coding }, then such a packet can be either $X_i^*$ or $Y_j^*$, see Table~\ref{tab:reactive}. 

\par We have two sub-cases in this scenario. Case 4.1: If the packet is $X_i^*$, we pass such an $X_i^*$ to the upper layer. Then $d_1$ examines whether it has stored $[X_i^*+Y_j^*]$ in its buffer. If so, use $X_i^*$ to decode $Y_j^*$ and insert $Y_j^*$ to the buffer. If not, store a separate copy of $X_i^*$ in the buffer even though one copy of $X_i^*$ has already been passed to the upper layer. Case 4.2: If the packet is $Y_j^*$, then by Table~\ref{tab:reactive} $d_1$ must have received the linear sum $[X_i^*+Y_j^*]$ in the corresponding {\sc Premixing} operation in the past. Therefore, $[X_i^*+Y_j^*]$ must be in the buffer of $d_1$ already. We can thus use $Y_j^*$ and $[X_i^*+Y_j^*]$ to decode the desired $X_i^*$. Receiver $d_1$ then passes the decoded $X_i^*$ to the upper layer and stores $Y_j^*$ in its buffer.

\par Case 5: If the received packet is generated by {\sc Degenerate XOR-1}, then such a packet can be either $X_i$ or $Y_j$, where $Y_j$ are those packets in $Q^1_{\{2\}}$ but coming from {\sc Reactive Coding}, see Fig.~\ref{fig:7-type}. Case 5.1: If the packet is $X_i$, we pass such an $X_i$ to the upper layer. Case 5.2: If the packet is $Y_j$, then from Table~\ref{tab:reactive}, it must be corresponding to the intersection of the row of $\rcpt=\overline{d_1}d_2$ and the column of $\rcpt^*=d_1\overline{d_2}$. As a result, $d_1$ must have received the corresponding $[X_i+Y_j]$ in the {\sc Premixing} operation. $d_1$ can thus use the received $Y_j$ to decode the desired $X_i$ and then pass $X_i$ to the upper layer.

\par Case 6: the received packet is generated by {\sc Degenerate XOR-2}. Consider two subcases. Case 6.1: the received packet is $X_i$. It is clear from Fig.~\ref{fig:7-type} that such $X_i$ must come from {\sc Reactive-Coding} since any packet from $Q_\emptyset^2$ to $Q_{\{1\}}^2$ must be a $Y_j$ packet. By Table~\ref{tab:reactive} and the row corresponding to $\rcpt=d_1\overline{d_2}$,  any $X_i\in Q^2_{\{1\}} $ that came from {\sc Reactive-Coding} must correspond to the column of $\rcpt^*=\overline{d_1}d_2$. By the second half of Case 4.1, such $X_i\in Q^2_{\{1\}}$ must be in the buffer of $d_1$ already. As a result, $d_1$ can simply ignore any $X_i$ packet it receives from {\sc Degenerate XOR-2}. Case 6.2: the received packet is $Y_j$. By the discussion of Case 2, if the $Y_j\in Q^2_{\{1\}}$ came from {\sc Non-Coding-2}, then it must be in the buffer of $d_1$ already. As a result, $d_1$ can simply ignore those $Y_j$ packets. If the $Y_j\in Q^2_{\{1\}}$ came from {\sc Reactive-Coding}, then by Table~\ref{tab:reactive} and the row corresponding to $\rcpt=d_1\overline{d_2}$, those $Y_j\in Q^2_{\{1\}}$ must correspond to the column of either $\rcpt^*=d_1\overline{d_2}$ or $\rcpt^*=d_1d_2$. By the first half of Case 4.1 and by Case 4.2, such $Y_j\in Q^2_{\{1\}}$ must be in the buffer of $d_1$ already. Again, $d_1$ can simply ignore those $Y_j$ packets. From the discussion of Cases 6.1 and 6.2, {\em any packet generated by {\sc Degenerate XOR-2} is already known to $d_1$, and nothing needs to be done in this case.}\footnote{Cases 5 and 6 echoes our previous arguments that any packet in $Q^2_{\{1\}}$ (which can be either $X_i$ or $Y_j$)  is information-equivalent to a session-2 packet that has been overheard by $d_1$.}

\par  Case 7: the received packet is generated by {\sc Classic-XOR}. Since we have shown in Case~6 that any packet in $Q^2_{\{1\}}$ is known to $d_1$, receiver $d_1$ can simply subtract the $Q^2_{\{1\}}$ packet from the linear sum received in Case 7. As a result, from $d_1$'s perspective, it is no different than directly receiving a $Q^1_{\{2\}}$ packet, i.e., Case 5. $d_1$ thus repeats the decoding  operation and buffer management in the same way as in Case 5.

{\bf Periodically pruning the memory:} In the above discussion, we elaborate which packets $d_1$ should store in its buffer and how to use them for decoding, while assuming no packet will ever be removed from the buffer. In the following, we discuss how to remove packets from the buffer of $d_1$.

\par We first notice that by the discussion of Cases 1 to 7, the uncoded packets in the buffer of $d_1$, i.e., those of the form of either $X_i$ or $Y_j$, are used  for decoding  {\em only  in the scenario of Case 7}. Namely, they are used to remove the $Q^2_{\{1\}}$ packet participating in the linear sum of {\sc Classic-XOR}. As a result, periodically we let $s$ send to $d_1$ the list 
of all packets in $Q^2_{\{1\}}$. After receiving the list, $d_1$ simply removes from its buffer any uncoded packets $X_i$ and/or $Y_j$ that are no longer in $Q^2_{\{1\}}$.

\par We then notice that by the discussion of Cases 1 to 7, the linear sum $[X_i+Y_j]$ in the buffer of $d_1$ is only used in one of the following two scenarios: (i) To decode $Y_j$ in Case 4.1 or to decode $X_i$ in Case 4.2; and (ii) To decode $X_i$ in Case 5.2. As a result, the $[X_i+Y_j]$ in the buffer is ``useful" only if one of the following two conditions are satisfied: (a) The corresponding tuple $(\rcpt,X_i,Y_j)$ is still in $Q_\text{mix}$, which corresponds to the scenarios of Cases 4.1 and 4.2; and (b) If the participating $Y_j$ is still in $Q^1_{\{2\}}$. By the above observation, periodically we let $s$ send to $d_1$ the list of all packets in $Q^1_{\{2\}}$ and $Q_\text{mix}$. After receiving the list, $d_1$ simply removes from its buffer any linear sum $[X_i+Y_j]$ that satisfies neither (a) nor (b).

The above pruning mechanism ensures that only the packets useful for future decoding are kept in the buffer of $d_1$ and $d_2$. Furthermore, it also leads to the following lemma.
\begin{lemma}\label{lemma:bound-receiver-buffer}Assume the lists of packets in $Q^1_{\{2\}}$, $Q^2_{\{1\}}$, and $Q_\text{mix}$ are sent to $d_1$ after every time slot. The number of packets in the buffer of $d_1$ is upper bounded by $|Q^1_{\{2\}}|+|Q^2_{\{1\}}|+|Q_\text{mix}|$.
\end{lemma}

The proof of Lemma~\ref{lemma:bound-receiver-buffer} is provided in \cite{KuoWang14:techrepstdsub}.

Lemma~\ref{lemma:bound-receiver-buffer} implies that as long as the queues in the vr-network are stabilized, the actual memory usage at both the source and the destinations can be stabilized simultaneously. 

{\em Remark:} 
Each transmitted packet is either an uncoded packet or a binary-XOR of two packets. Therefore, during transmission we only need to store 1 or 2 packet sequence numbers in the header of the uncoded/coded packet, depending on whether we send an uncoded packet or a linear sum. \change{The overhead of updating the packet list is omitted but we can choose only to update it periodically.} The communication overhead of the proposed scheme is thus small.

\section{The Proposed Scheduling Solution \label{sec:new_scheduling}}

In this section, we aim to solve Challenge 2 in Section~\ref{sec:introduction}. The main tool that we use to stabilize the vr-network is stochastic processing networks (SPNs). In the following, we will discuss the basic definitions, existing results on a special class of SPNs, and our throughput-optimal scheduling solution.

\subsection{The Main Features of SPNs}
The SPN is a generalization of the store-and-forward networks. In an SPN, a packet cannot be transmitted directly from one queue to another queue through links. Instead, it must first be processed by a unit called ``Service Activity'' (SA). The SA first collects a certain amount of packets from one or more queues (named the {\em input queues}), jointly processes/consumes these packets, generates a new set of packets, and finally redistributes them to another set of queues (named the {\em output queues}). The number of consumed packets may be different than the number of generated packets. There is one critical rule: {\em An SA can be activated only when all its input queues can provide enough amount of packets for the SA to process.} This rule captures directly the INC behavior and thus makes INC a natural application of SPNs. Other applications of SPNs include the video streaming 
and  Map-\&-Reduce scheduling. 

All the existing SPN scheduling solutions \cite{jiang2009stable,huang2011utility} assume a special class of SPNs, which we call SPNs with deterministic departure, which is quite  different from our INC-based vr-network. The reason is as follows. When a packet is broadcast by $s$, it can arrive at a random subset of receivers. Therefore, the vr-packets move among the vr-queues according to some probability distribution. We call the SPN model that allows random departure service rates ``the SPN with random departure.'' It turns out that random departure presents a unique challenge for SPN scheduling. See \cite{jiang2009stable} for an example of such a challenge and also see the discussion in \cite{KuoWang14:techrepstdsub}.

\subsection{A Simple SPN Model with Random Departure} \label{subsec:01SPN}
We now formally define a random SPN model that includes the INC vr-network in Section~\ref{sec:new-coding} as a special example. Consider a time-slotted system with i.i.d.\ channel quality $\cq(t)$. A (0,1) random SPN consists of three components: the input activities (IAs), the service activities (SAs), and the queues. Suppose that there are $K$ queues, $M$  IAs, and $N$ SAs.

{\bf Input Activities:} Each IA represents a session (or a flow) of packets. Specifically, when an IA $m$ is activated, it injects a deterministic number of $\alpha_{k,m}$ packets to queue $k$ where $\alpha_{k,m}$ is of integer value. We use $\im\in \RR^{K\times M}$ to denote the ``input matrix" with the $(k,m)$-th entry equals to $\alpha_{k,m}$, for all $m$ and $k$. At each time $t$, a random subset of IAs will be activated. Equivalently, we define $\av\stdef(a_1(t),a_2(t),\cdots, a_M(t))\in \{0,1\}^M$ as the random ``arrival vector" at time $t$. If $\avm{m}=1$, then IA $m$ is activated at time $t$. We assume that the random vector  $\av$ is i.i.d.\ over time with the average rate vector $\ve{R} = \EE\{\av\}$. In our setting, the $\im$ matrix is a fixed (deterministic) system parameter and all the randomness of IAs lies in $\av$.

{\bf Service Activities:} For each service activity SA $n$, we define the {\em input queues} of SA $n$ as the queues which are required to provide some packets when SA $n$ is activated. Let $\SAin{n}$ denote the collection of the input queues of SA $n$. Similarly, we define the {\em output queues} of SA $n$ as the queues which will possibly receive packets when SA $n$ is activated, and let $\SAout{n}$ be the collection of the output queues of SA $n$. I.e., when SA $n$ is activated, it consumes packets from queues in $\SAin{n}$, and generates new packets and sends them to queues in $\SAout{n}$. We assume that $\cq(t)$ does not change $\SAin{n}$ and $\SAout{n}$.

\par There are 3 SA-activation rules in a (0,1) random SPN:

\par {\em SA-Activation Rule 1:} SA $n$ can be activated only if for all $k\in\SAin{n}$, queue $k$ has at least 1 packet in the queue. For future reference, we say SA $n$ is {\em feasible} at time $t$ if at time $t$ queue $k$ has at least 1 packet for all $k\in\SAin{n}$. Otherwise, we say SA $n$ is infeasible at time $t$.

{\em SA-Activation Rule 2:} When SA $n$ is activated with the channel quality $c$ (assuming SA $n$ is feasible), the number of packets leaving queue $k$ is a binary random variable, $\bin{k}{n}(c)$, with mean $\overline{\bin{k}{n}(c)}$ for all $k\in\SAin{n}$.

\par Note that there is a subtlety in Rules 1 and 2. By Rule 2, when we activate an SA $n$, it sometimes consumes zero packet from its input queues. However, even if it may consume zero packet, Rule 1 imposes that all input queues must always have at least 1 packet before we can activate an SA. Such a subtlety is important for our vr-network. For example, we can schedule {\sc Premixing} in Fig.~\ref{fig:7-type} only when both $Q^1_\emptyset$ and $Q^2_\emptyset$ are non-empty. But whether {\sc Premixing} actually consumes any $Q^1_\emptyset$ and $Q^2_\emptyset$ packets depending on the random reception event of the transmission.

{\em SA-Activation Rule 3:} When SA $n$ is activated with the channel quality $c$ (assuming SA $n$ is feasible), the number of packets entering queue $k$ is a binary random variable, $\bout{k}{n}(c)$, with mean $\overline{\bout{k}{n}(c)}$ for all $k\in\SAout{n}$.

Let $\Bin(c)\in\RR^{K*N}$ be the {\em random input service matrix} under channel quality $c$ with the $(k,n)$-entry equals to $\bin{k}{n}(c)$, and let $\Bout(c)\in\RR^{K*N}$ be the {\em random output service matrix} under channel quality $c$ with the $(k,n)$-entry equals to $\bout{k}{n}(c)$. The expectations of $\Bin(c)$ and $\Bout(c)$ are denoted by $\overline{\Bin(c)}$ and $\overline{\Bout(c)}$, respectively. We assume that given any channel quality $c\in\bigcq$, both the input and output service matrix $\Bin(c)$ and $\Bout(c)$ are independently distributed over time.

{\bf Scheduling of the SAs:} At the beginning of each time $t$, the SPN scheduler is made aware of the current channel quality $\cq(t)$ and can choose to ``activate'' at most one 
SA.  Let $\sv\in \{0,1\}^N$ be the ``service vector" at time $t$. If the $n$-th coordinate $\svm{n}=1$, then it implies that we choose to activate SA $n$ at time $t$. Let $\mathfrak{X}$ denote the set of vectors that contains all Dirac delta vectors and the all-zero vector, i.e., those vectors that can be activated at any given time slot. Define $\Lambda$ to be the convex hull of $\mathfrak{X}$ and let $\Lambda^\circ$ be the interior of $\Lambda$.

{\bf Other Technical Assumptions:}  We also use the following 2 technical assumptions. Assumption 1: The input/output queues $\SAin{n}$ and $\SAout{n}$ of the SAs can be used to plot the corresponding SPN. We assume that the corresponding SPN is acyclic. Assumption 2: For any $\cq(t)=c$, the expectation of $\bin{k}{n}(c)$ (resp. $\bout{k}{n}(c)$) with $k\in\SAin{n}$ (resp. $k\in\SAout{n}$) is in $(0,1]$. Assumptions 1 and 2 are used to rigorously prove the mean-rate stability region, which eliminate, respectively, the cyclic setting and the limiting case in which the Bernoulli random variables are always 0.

\par One can easily verify that the above (0,1) random SPN model includes the vr-network in Fig.~\ref{fig:7-type} as a special example.

\subsection{The Proposed Scheduler For (0,1) Random SPNs} \label{subsec:proposed-DMW}
We borrow the wisdom of deficit maximum weight (DMW) scheduling \cite{jiang2009stable}. Specifically, our scheduler maintains a real-valued counter $q_k(t)$, called the virtual queue length, for each queue $k$. Initially, $q_k(1)$ is set to 0. For comparison, the actual queue length is denoted by $Q_k(t)$.

The key feature of the scheduler is that it makes its decision based on $q_k(t)$ instead of $Q_k(t)$. Specifically, for each time $t$, we compute the ``preferred\footnote{Sometimes we may not be able to execute/schedule the preferred service activities chosen by \eqref{eq:DMW-schedule}. This is the reason why we only call the $\x^*(t)$ vector in \eqref{eq:DMW-schedule} a preferred choice, instead of a scheduling choice.} service vector"  by
\begin{align}
&\ve{x}^*(t) = \arg\max_{\ve{x}\in \mathfrak{X}} \ve{d}^\tran(t)\cdot \ve{x}, \label{eq:DMW-schedule}\\
\text{where}\quad\quad&\ve{d}(t)=\left(\overline{\Bin(\cq(t))}-\overline{\Bout(\cq(t))}\right)^\tran \ve{q}(t)\label{eq:new-backpressure}
\end{align}
is the back pressure vector; $\ve{q}(t)$ is the vector of the virtual queue lengths; and we recall that the notations $\overline{\Bin(\cq(t))}$ and $\overline{\Bout(\cq(t))}$ are the expectations when the channel quality $\cq(t)=c$. Since we assume that each vector in $\mathfrak{X}$ has at most 1 non-zero coordinate, \eqref{eq:DMW-schedule} and \eqref{eq:new-backpressure} basically find the preferred SA $n^*$ in time $t$. We then check whether the preferred SA $n^*$ is feasible. If so, we officially schedule SA $n^*$. If not, we let the system to be idle,\footnote{The reason of letting the system idle is to facilitate rigorous stability analysis. In practice, when the preferred choice is infeasible, we can choose a feasible SA $n$ with the largest back-pressure computed by the actual queue lengths $Q_k(t)$ instead of the virtual queue lengths $q_k(t)$.} i.e., the actually scheduled service vector $\ve{x}(t)=\0$ is now all-zero.

\par Regardless of whether the preferred SA $n^*$ is feasible or not, we update $\ve{q}(t)$ by:
\begin{align}
\ve{q}(t+1) =& \ve{q}(t)+\im\cdot \ve{a}(t)\nonumber \\
 &+ \left(\overline{\Bout(\cq(t))}-\overline{\Bin(\cq(t))}\right)\cdot \ve{x}^*(t). \label{eq:DMW-update-new}
\end{align}
Note that the actual queue length $Q_k(t)$ is updated in a way very different from  \eqref{eq:DMW-update-new}. 
If the preferred SA $n^*$ is not feasible, then the system remains idle and $Q_k(t)$ changes if and only if there is any new packet arrival. If SA $n^*$ is feasible, then $Q_k(t)$ is updated based on the actual packet movement. While the actual queue lengths $Q_k(t)$ is always $\geq 0$, the virtual queue length $\ve{q}(t)$ can be strictly negative when updated via \eqref{eq:DMW-update-new}.

The above scheduling scheme is denoted by $\SCH_\text{avg}$ since 
\eqref{eq:DMW-update-new} is based on the average departure rate.

\subsection{Performance Analysis\label{subsec:analysis}}
The following two propositions characterize the mean-rate stability region of any (0,1) random SPN. 
\begin{proposition}\label{prop:outer}
A rate vector $\ve{R}$ can be mean-rate stabilized only if
there exist $\ve{s}_c\in \Lambda$ for all $c\in\bigcq$ such that
\begin{align}
\im\cdot \ve{R}+\sum_{c\in\bigcq}f_c\cdot \overline{\Bout(c)}\cdot \ve{s}_c=\sum_{c\in\bigcq}f_c\cdot \overline{\Bin(c)}\cdot \ve{s}_c.\label{eq:new-balance}
\end{align}
\end{proposition}

Proposition~\ref{prop:outer} can be derived by conventional flow conservation arguments as in \cite{jiang2009stable} and the proof is thus omitted.

\begin{proposition}\label{prop:inner}
For 
any rate vector $\ve{R}$, if there exist $\ve{s}_c\in \Lambda^\circ$ for all $c\in\bigcq$ such that \eqref{eq:new-balance} holds, then the proposed scheme $\SCH_\text{avg}$ in Section~\ref{subsec:proposed-DMW} can mean-rate stabilize the (0,1) random SPN with arrival rate $\ve{R}$.
\end{proposition}

{\em Outline of the proof of Proposition~\ref{prop:inner}: }Let each queue $k$ keep another two real-valued counters $\qkinter{k}(t)$ and $\Qkinter{k}(t)$, termed the {\em intermediate virtual queue length} and {\em intermediate actual queue length}. 
There are thus 4 different queue length values\footnote{$\qkinter{k}(t)$ and $\Qkinter{k}(t)$ are used only for the proof and are not needed when running the scheduling algorithm.} $q_k(t)$, $\qkinter{k}(t)$, $\Qkinter{k}(t)$, and $Q_k(t)$ for each queue $k$. To prove $\ve{Q}(t)$, the vector of actual queue lengths, can be stabilized, we will show that both $\Qkinter{k}(t)$ and $|Q_k(t)-\Qkinter{k}(t)|$ can be mean-rate stabilized by $\SCH_\text{avg}$ for all $k$. Since the summation of mean-rate stable random processes is still mean-rate stable, $\ve{Q}(t)$ can thus be mean-rate stabilized by $\SCH_\text{avg}$. 

With the above road map, we now specify the update rules for $\qkinter{k}(t)$ and $\Qkinter{k}(t)$. Initially, $\qkinter{k}(1)$ and $\Qkinter{k}(1)$ are set to 0 for all $k$. In the end of each time $t$, we compute $\qinter(t+1)$ using the preferred schedule $\ve{x}^*(t)$ chosen by $\SCH_{\text{avg}}$:
\begin{align}
\qinter(t+1) =& \qinter(t)+\im\cdot \ve{a}(t)\nonumber \\
&+ \left(\Bout(\cq(t))-\Bin(\cq(t))\right)\cdot \ve{x}^*(t). \label{eq:DMW-update-inter}
\end{align}
Comparing \eqref{eq:DMW-update-inter} and 
\eqref{eq:DMW-update-new}, we can see that $\ve{q}^\text{inter}(t)$ is updated by the {\em realization} of the input/output service matrices while $\ve{q}(t)$ is updated by the {\em expected} input/output service matrices.

We can rewrite \eqref{eq:DMW-update-inter} in the following equivalent form:
\begin{align}
\qkinter{k}(t+1)=\qkinter{k}(t)-\mu_{\text{out},k}(t)+\mu_{\text{in},k}(t),~\forall k, \label{eq:qk-inter-update}
\end{align}
where
\begin{align}
&\mu_{\text{out},k}(t) = \sum_{n=1}^N \left(\bine{k}{n}(\cq(t)) \cdot x_n^*(t) \right), \label{eq:mu_out_def} \\
&\mu_{\text{in},k}(t) = \sum_{m=1}^M\left( \alpha_{k,m}\cdot a_m(t) \right)+\sum_{n=1}^N \left(\bout{k}{n}(\cq(t))\cdot x_n^*(t)\right). \label{eq:mu_in_def}
\end{align}
Here, $\mu_{\text{out},k}$ is the amount of packets coming ``out of queue $k$", which is decided by the ``input rates of SA $n$". Similarly,
$\mu_{\text{in},k}$ is the amount of packets ``entering queue $k$", which is decided by the ``output rates of SA $n$" and the packet arrival rates. We now update $\Qinter(t+1)$ by
\begin{align}
\Qkinter{k}(t+1)=\left(\Qkinter{k}(t)-\mu_{\text{out},k}(t)\right)^+ + \mu_{\text{in},k}(t),~\forall k, \label{eq:Qk-inter-update}
\end{align}
where $(v)^+ = \max\{0,v\}$.

The difference between $\qkinter{k}(t)$ and $\Qkinter{k}(t)$ is that the former can be still be strictly negative when updated via \eqref{eq:qk-inter-update} while we enforce the latter to be non-negative.

\par To compare $\Qkinter{k}(t)$ and $Q_k(t)$, we observe that by \eqref{eq:Qk-inter-update}, $\Qkinter{k}(t)$ is updated by the preferred service vector $\ve{x}^*(t)$ without considering whether the preferred SA $n^*$ is feasible or not. In contrast, the update rule of the actual queue length $Q_k(t)$ is quite different. For example, if SA $n^*$ is infeasible, then the system remains idle and we have
\begin{align}
Q_k(t+1)=Q_k(t) + \sum_{m=1}^M\left( \alpha_{k,m}\cdot a_m(t) \right).\label{eq:new-actual}
\end{align}
Note that \eqref{eq:new-actual} differs significantly from \eqref{eq:Qk-inter-update}. For example, say we have $Q_k(t)=0$ to begin with. When SA $n^*$ is infeasible, by \eqref{eq:new-actual} the aggregate increase of $Q_k(t)$ depends only on the new packet arrivals. But the aggregate increase of $\Qkinter{k}(t)$, assuming $\Qkinter{k}(t)=0$, depends on the service rates of the preferred $x_n^*(t)$ as well,\footnote{In the original DMW algorithm \cite{jiang2009stable}, the quantity ``actual queue length'' is updated by \eqref{eq:Qk-inter-update} instead of \eqref{eq:new-actual}. The ``actual queue lengths in \cite{jiang2009stable}'' thus refer to the register value $\Qkinter{k}(t)$ rather than the number of physical packets in the buffer/queue. In this work, we rectify this inconsistency by renaming ``the actual queue lengths in \cite{jiang2009stable}'' the ``intermediate actual queue lengths $\Qkinter{k}(t)$.''}
 see the two terms in \eqref{eq:mu_in_def}.

We first focus on the absolute difference $|Q_k(t)-\Qkinter{k}(t)|$. 
We use $n(t)$ to denote the preferred SA suggested by the back-pressure scheduler in \eqref{eq:DMW-schedule} and \eqref{eq:new-backpressure}. We now define an event, which is called the {\em null activity} of queue $k$ at time $t$. We say the null activity occurs at queue $k$ if (i) $k\in\SAin{n(t)}$ and (ii) $\Qkinter{k}(t)<\bine{k}{n(t)}(\cq(t))$. That is, the null activity describes the event that the preferred SA shall consume the packets in queue $k$ (since $k\in \SAin{n(t)}$) but at the same time $\Qkinter{k}(t)<\bine{k}{n}(\cq(t))$. Note that the null activity is defined based on comparing the intermediate actual queue length $\Qkinter{k}(t)$ and the actual realization of the packet consumption $\bine{k}{n(t)}(\cq(t))$. For comparison, whether the SA $n(t)$ is feasible depends on whether the actual queue length $Q_k(t)$ is larger or less than 1. Therefore the null activities are not directly related to the event that SA $n(t)$ is infeasible.\footnote{If $\Qkinter{k}(t)\geq Q_k(t)$ for all $k$ and $t$ with probability 1, then the event ``SA $n(t)$ is infeasible" implies the null activity for at least one of the input queues of SA $n(t)$. One can then upper bound the frequency of SA $n(t)$ being infeasible by upper bounding how frequently we encounter the null activities of queue $k$  as suggested in \cite{jiang2009stable}. Unfortunately, we have proven that $\Qkinter{k}(t)< Q_k(t)$ with strictly positive probability for some $k$ and $t$. The arguments in \cite{jiang2009stable} thus do not hold. 
Instead, we introduce a new {\em expectation-based} dominance relationship in Lemma~\ref{lemma:bound-Qk} and use it to establish the connection between null activities and the instants SA $n(t)$ is infeasible. Also see \cite{KuoWang14:techrepstdsub}.}  

\par Let $N_{\mathsf{NA},k}(t)$ be the aggregate number of null activities occurred at queue $k$ up to time $t$. That is, 
\begin{align}
\Nna{k}(t)\eqdef 
&\sum_{\tau=1}^{t} I(k\in\SAin{n(\tau)}) \cdot  I(\Qkinter{k}(\tau)<\bine{k}{n(\tau)}(\cq(\tau))) \nonumber
\end{align}
where $I(\cdot)$ is the indicator function.
We then have


\begin{lemma}\label{lemma:bound-Qk}
For all $k=1,2,\cdots,K$, there exist $K$ non-negative coefficients $\gamma_{1},...,\gamma_{K}$ such that
\begin{align}
&\EE\left(|Q_k(t)-\Qkinter{k}(t)|\right) \leq \sum_{\tilde{k}=1}^K \gamma_{\tilde{k}} \EE\left(\Nna{\tilde{k}}(t)\right). \label{eq:bound-Qk}
\end{align}
for all $t=1$ to $\infty$.
\end{lemma}

The proof of Lemma~\ref{lemma:bound-Qk} is relegated to Appendix~A of \cite{KuoWang14:techrepstdsub}. In Appendix~D of \cite{KuoWang14:techrepstdsub}, we prove that  $\Qkinter{k}(t)$ and $\Nna{k}(t)$ can be mean-rate stabilized by $\SCH_\text{avg}$ for all $k$. Therefore, by Lemma~\ref{lemma:bound-Qk},  $|Q_k(t)-\Qkinter{k}(t)|$ can be mean-rate stabilized and so can $Q_k(t)$. Proposition~\ref{prop:inner} is thus proven.

\section{The Combined Dynamic INC Solution} \label{sec:combined}

We now combine the discussions in Sections~\ref{sec:new-coding} and \ref{sec:new_scheduling}. As discussed in Section~\ref{sec:new-coding}, the 7 INC operations form a vr-network as described in Fig.~\ref{fig:7-type}. How $s$ generates an NC packet is now converted to a scheduling problem of the vr-network of Fig.~\ref{fig:7-type}, which has $K=5$ queues, $M=2$ IAs, and $N=7$ SAs. The 5-by-2 input matrix $\mathcal{A}$ contains 2 ones, since the packets arrive at either $Q^1_\emptyset$ or $Q^2_\emptyset$. Given the channel quality $\cq(t)=c$, the expected input / output service matrices $\avg{\Bin(c)}$ and $\avg{\Bout(c)}$ can be derived from Table~\ref{tab:transition}.

\par For illustration, suppose that $\cq(t)$ is Bernoulli with parameter $1/2$  (i.e., flipping a perfect coin and the relative frequency $f_0=f_1=0.5$). Also suppose that when $\cq(t)=0$, with probability $0.5$ (resp.\ $0.7$) $d_1$ (resp.\ $d_2$) can successfully receive a packet transmitted by $s$; and when $\cq(t)=1$, with probability $2/3$ (resp.\ $1/3$) $d_1$ (resp.\ $d_2$) can successfully receive a packet transmitted by $s$. Further assume that all the success events of $d_1$ and $d_2$ are independent. If we order the 5 queues as $\left[\ve{Q}_\emptyset^1,\ve{Q}_\emptyset^2,\ve{Q}_{\{2\}}^1,\ve{Q}_{\{1\}}^2,\ve{Q}_\text{mix}\right]$, the 7 service activities as $\left[\text{NC1},\text{NC2},\text{DX1},\text{DX2},\text{PM},\text{RC},\text{CX}\right]$, then the matrices of the SPN become
\begin{align}
&\im = \left[
           \begin{array}{ccccc}
             1 & 0 & 0 & 0 & 0 \\
             0 & 1 & 0 & 0 & 0 \\
           \end{array}
         \right]^\tran, \nonumber \\
&\overline{\Bin(0)} =\left[
        \begin{array}{ccccccc}
          0.85 & 0 & 0 & 0 & 0.85 & 0 & 0 \\
          0 & 0.85 & 0 & 0 & 0.85 & 0 & 0 \\
          0 & 0 & 0.5 & 0 & 0 & 0 & 0.5 \\
          0 & 0 & 0 & 0.7 & 0 & 0 & 0.7 \\
          0 & 0 & 0 & 0 &  0 &  0.85 & 0 \\
        \end{array}
      \right], \nonumber\\
&\overline{\Bin(1)} =\left[
         \begin{array}{ccccccc}
           7/9 & 0 & 0 & 0 & 7/9 & 0 & 0 \\
           0 & 7/9 & 0 & 0 & 7/9 & 0 & 0 \\
           0 & 0 & 2/3 & 0 & 0 & 0 & 2/3 \\
           0 & 0 & 0 & 1/3 & 0 & 0 & 1/3 \\
           0 & 0 & 0 & 0 &  0 &  7/9 & 0 \\
         \end{array}
       \right], \nonumber\\
&\overline{\Bout(0)} = \left[
        \begin{array}{ccccccc}
          0 & 0 & 0 & 0 & 0 & 0 & 0 \\
          0 & 0 & 0 & 0 & 0 & 0 & 0 \\
          0.35 & 0 & 0 & 0 & 0 & 0.35 & 0 \\
          0 & 0.15 & 0 & 0 & 0 & 0.15 & 0 \\
          0 & 0 & 0 & 0 &  0.85 &  0 & 0 \\
        \end{array}
      \right], \nonumber\\
&\overline{\Bout(1)} = \left[
        \begin{array}{ccccccc}
          0 & 0 & 0 & 0 & 0 & 0 & 0 \\
          0 & 0 & 0 & 0 & 0 & 0 & 0 \\
          1/9 & 0 & 0 & 0 & 0 & 1/9 & 0 \\
          0 & 4/9 & 0 & 0 & 0 & 4/9 & 0 \\
          0 & 0 & 0 & 0 &  7/9 &  0 & 0 \\
        \end{array}
      \right]. \nonumber 
\end{align}
For example, the seventh column of $\overline{\Bin(0)}$ indicates that when $\cq(t)=0$ and {\sc Classic-XOR} is activated, with probability 0.5 (resp.\  0.7) 1 packet will be consumed from queue $\ve{Q}_{\{2\}}^1$ (resp.\ $\ve{Q}_{\{1\}}^2$). The third row of $\overline{\Bout(1)}$ indicates that when $\cq(t)=1$,  queue $\ve{Q}_{\{2\}}^1$ will increase by 1 with probability 1/9 (resp.\ 1/9) if {\sc Non-Coding-1} (resp.\ {\sc Reactive-Coding}) is activated since it corresponds to the event that $d_1$ receives the transmitted packet but $d_2$ does not.

\par We can now use the proposed DMW scheduler in \eqref{eq:DMW-schedule}, \eqref{eq:new-backpressure}, and \eqref{eq:DMW-update-new} to compute the preferred scheduling decision in every time $t$. We activate the preferred decision if it is feasible. If not, then the system remains idle.

\par For general channel parameters (including but not limited to this simple example), after computing the $\overline{\Bin(c)}$ and $\overline{\Bout(c)}$ of the vr-network in Fig.~\ref{fig:7-type} with the help of Table~\ref{tab:transition}, we can explicitly compare the mean-rate stability region in Propositions~\ref{prop:outer} and \ref{prop:inner} with the Shannon capacity region in \cite{WangHan14}. In the end, we have the following proposition.

\begin{proposition}\label{prop:combine-capacity} The mean-rate stability region of the proposed INC-plus-SPN-scheduling scheme always matches the block-code capacity of time-varying channels.
\end{proposition}
\noindent
A detailed proof of Proposition~\ref{prop:combine-capacity} is provided in Appendix~E of the technical report \cite{KuoWang14:techrepstdsub}.

{\em Remark:} During numerical simulations, we notice that we 
can further revise the proposed scheme to reduce the actual queue lengths $Q_k(t)$ by $\approx 50\%$ even though we do not have any rigorous proofs/performance guarantees for the revised scheme. That is, when making the scheduling decision by \eqref{eq:DMW-schedule}, we can compute $\ve{d}(t)$ by
\begin{align}
\ve{d}(t)=\left(\overline{\Bin(\cq(t))}-\overline{\Bout(\cq(t))}\right)^\tran \ve{q}^\text{inter}(t)\label{eq:inter-backpressure}
\end{align}
where $\ve{q}^\text{inter}(t)$ is the intermediate virtual queue length defined in \eqref{eq:qk-inter-update}. The intuition behind is that the new back-pressure in \eqref{eq:inter-backpressure} allows the scheme to directly control $q_k^\text{inter}(t)$, which, when compared to the virtual queue $\ve{q}(t)$ in \eqref{eq:DMW-update-new}, is more closely related to the actual queue length\footnote{There are four types of queue lengths in this work: $\ve{q}(t)$, $\qinter(t)$, $\Qinter(t)$, and $\ve{Q}(t)$ and they range from the most artificially-derived $\ve{q}(t)$ to the most realistic metric, the actual queue length $\ve{Q}(t)$.} $Q_k(t)$.

\subsection{Extensions For Rate Adaption} \label{subsec:extention_ACM}
The proposed dynamic INC solution can be generalized for rate adaptation, also known as adaptive coding and modulation. For illustration, we consider the following example.

\par Consider 2 possible error correcting rates (1/2 and 3/4); 2 possible modulation schemes QPSK and 16QAM; and jointly there are 4 possible combinations. The lowest throughput combination is rate-1/2 plus QPSK and the highest throughput combination is rate-3/4 plus 16QAM. Assuming the packet size is fixed. If the highest throughput combination takes 1-unit time to finish sending 1 packet, then the lowest throughput combination will take 3-unit time. For these 4 possible (rate,modulation) combinations, we denote the unit-time to finish transmitting 1 packet as $T_1$ to $T_4$, respectively.

\par For the $i$-th (rate,modulation) combination, $i=1$ to $4$, source $s$ can measure the probability that $d_1$ and/or $d_2$ successfully hears the transmission, and denote the corresponding probability vector by $\vec{p}^{(i)}$. Source $s$ then uses $\vec{p}^{(i)}$ to compute the  $\overline{\Bini(c)}$ and $\overline{\Bouti(c)}$ for the vr-network when $\cq(t)=c$. At any time $t$, after observing $\cq(t)$ source $s$ computes the back-pressure by
\begin{align}
\ve{d}^{(i)}(t)=\left(\overline{\Bini(\cq(t))}-\overline{\Bouti(\cq(t))} \right)^\tran \ve{q}(t). \nonumber
\end{align}
We can now compute the preferred scheduling choice by
\begin{align}
\argmax_{i\in\{1,2,3,4\},\ve{x}\in\mathfrak{X}} \frac{\ve{d}^{(i)}(t)^\tran \cdot \ve{x} }{T_i}\label{eq:new-sch-ACM}
\end{align}
and update the virtual queue length $\ve{q}(t)$ by \eqref{eq:DMW-update-new}. Namely, the back-pressure $\ve{d}^{(i)}(t)^\tran \cdot \ve{x}$ is scaled inverse proportionally with respect to $T_i$, the time it takes to finish the transmission of 1 packet. If the preferred SA $n^*$ is feasible, then we use the $i^*$-th (rate,modulation) combination plus the coding choice $n^*$ for the current transmission. If the preferred SA $n^*$ is infeasible, then we let the system remain idle.

\par One can see that the new scheduler \eqref{eq:new-sch-ACM} automatically balances the packet reception status (the $\ve{q}(t)$ terms), the success overhearing probability of different (rate,modulation) (the $\overline{\Bini(\cq(t))}$ and $\overline{\Bouti(\cq(t))}$ terms), and different amount of time it takes to finish transmission of a coded/uncoded packet  (the $T_i$ term). In all the numerical experiments we have performed, the new scheduler \eqref{eq:new-sch-ACM} robustly achieves the optimal throughput with adaptive coding and modulation.

\subsection{Practical Issues}
In addition to the theoretic focus of this work, here we discuss two practical issues of the proposed solution.

{\em Delayed Feedback:} In this work, we assume that the ACK feedback is transmitted via a separate, error-free control channel {\em immediately after} each forward packet transmission. The error-free assumption is justified by the fact that in practice, ACK is usually transmitted through the lowest MCS level to ensure the most reliable transmission.

On the other hand, the instant feedback assumption may not hold in practice since delayed feedback mechanisms are used widely in real-world systems in order to minimize the number of transmission-reception transition intervals. For example, the mandatory Block-ACK mechanism in IEEE 802.11n standard forces the feedbacks to be aggregated and to be transmitted at the end of each Transmit Opportunity (TXOP) at once, instead of after reception of each packet.

\par Our proposed solution can be modified to accommodate the delayed feedback by incorporating the designs in \cite{LiWangLinJSAC11}. The main idea of \cite{LiWangLinJSAC11} is to pipelining the operations and let the base-station only processes those {\em properly acknowledged} packets. This converts the delayed feedback scenario to an equivalent instant feedback set-up. See \cite{LiWangLinJSAC11} for detailed discussion on handling delayed feedback.

{\em Scability: }Although the discussion of this work focuses exclusively on the 2-client case, there are many possible ways of extending the solution to more-than-2-client applications.\footnote{Unfortunately, we can no longer guarantee the optimality of the dynamic INC solution. This is because even the block-code capacity (Shannon capacity) for more than 3-client case remains largely unknown\cite{Wang10b}.} For example, the router of city-WiFi may serve multiple smart devices and laptops, and suppose we have 10 clients. Before transmission, we can first estimate how much throughput gain we can have if we group any two specific clients as a pair and perform INC on this pair. Then, we divide the clients into 5 pairs of clients that can lead to the highest throughput gain. After dividing the clients into pairs, we start the packet transmission and apply the dynamic INC solution within each pair. The detailed implementation of such an approach is beyond the scope of this work.

\section{Simulation Results\label{sec:simulation}}

\begin{figure}
\centering
  \includegraphics[width=9cm]{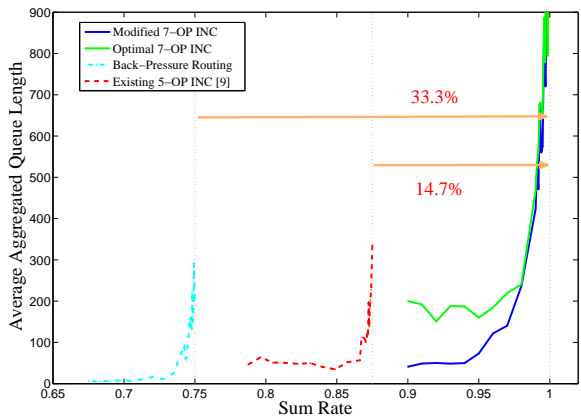}\\
  \caption{The backlog of four different schemes for a time-varying channel with $\cq(t)$ uniformly distributed on $\{1,2\}$, and the packet delivery probability being $\vec{p}=(0,0.5,0.5,0)$ if $\cq(t)=1$ and $\vec{p}=(0,0,0,1)$ if $\cq(t)=2$.}\label{fig:7vs5}
\end{figure}

We simulate the proposed optimal 7-operation INC + scheduling solution and compare the results  with the existing INC solutions and the (back-pressure) pure-routing solutions. \change{We use a custom built simulator in MATLAB. Even though we spent a quite amount of effort/pages on proving the correctness, the core algorithm is pretty simple and could be implemented in less than 200 lines of codes to compute the chosen coding operation at each iteration.}

In Fig.~\ref{fig:7vs5}, we simulate a simple time-varying channel situation first described in Section~\ref{subsubsec:degenerate-XOR}. Specifically, the channel quality $\cq(t)$ is i.i.d.\ distributed and for any $t$,  $\cq(t)$ is uniformly distributed on $\{1,2\}$. When $\cq(t)=1$, the success probabilities are $\vec{p}^{(1)}=(0,0.5,0.5,0)$ and when $\cq(t)=2$, the success probabilities are $\vec{p}^{(2)}=(0,0,0,1)$, respectively. We consider four different schemes: (i) Back-pressure (BP) + pure routing; (ii) BP + INC with 5 operations \cite{AthanasiadouGeorgiadis13}; (iii) The proposed DMW+INC with 7 operations, and (iv) The modified DMW+INC with 7 operations that use $\qkinter{k}(t)$ to compute the back pressure, see \eqref{eq:inter-backpressure}, instead of $q_k(t)$ in \eqref{eq:new-backpressure}.

\par We choose perfectly fair $(R_1,R_2)=(\theta, \theta)$ and gradually increase the $\theta$ value and plot the stability region. For each experiment, i.e., each $\theta$, we run the schemes for $10^5$ time slots. The horizontal axis is the sum rate $R_1+R_2=2\theta$ and the vertical axis is the aggregate backlog  (averaged over 10 trials) in the end of $10^5$ slots. By \cite{WangHan14}, the sum rate Shannon capacity is 1 packet/slot, the best possible rate for 5-OP INC is 0.875 packet/slot, and the best pure routing rate is 0.75 packet/slot, which are plotted as vertical lines in Fig.~\ref{fig:7vs5}.  The simulation results confirm our analysis. The proposed 7-operation dynamic INC has a stability region matching the Shannon block code capacity and provides $14.7\%$ throughput improvement over the 5-operation INC, and $33.3\%$ over the pure-routing solution.

\par Also, both our original proposed solution (using $q_k(t)$) and the modified solution (using $\qkinter{k}(t)$) can approach the stability region while the modified solution has smaller backlog. This phenomenon is observed throughout all our experiments. As a result, in the following experiments, we only report the results of the modified solution.

\begin{figure}
\centering
\subfigure[$(f_1,f_2,f_3,f_4)=(0.15,0.15,0.35,0.35)$. \label{fig:random}]{
  \includegraphics[width=9cm]{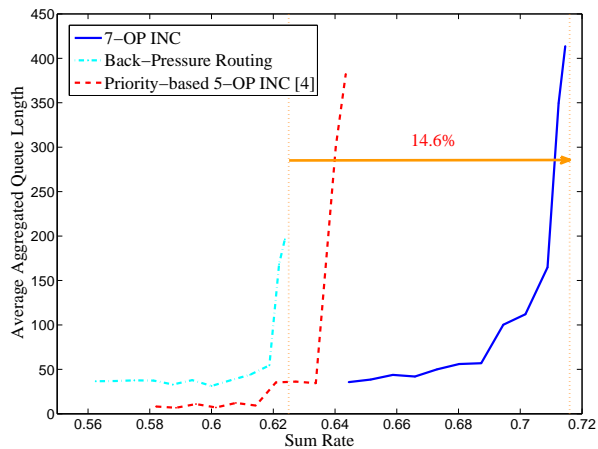}}
\subfigure[$(f_1,f_2,f_3,f_4)=(0.25,0.25,0.25,0.25)$. \label{fig:round}]{
  \includegraphics[width=9cm]{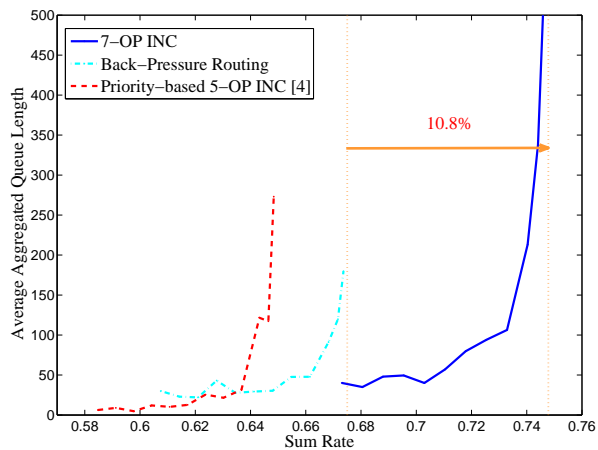}}
\subfigure[Periodic channel quality with 3-repetition. \label{fig:periodic}]{
	\includegraphics[width=9cm]{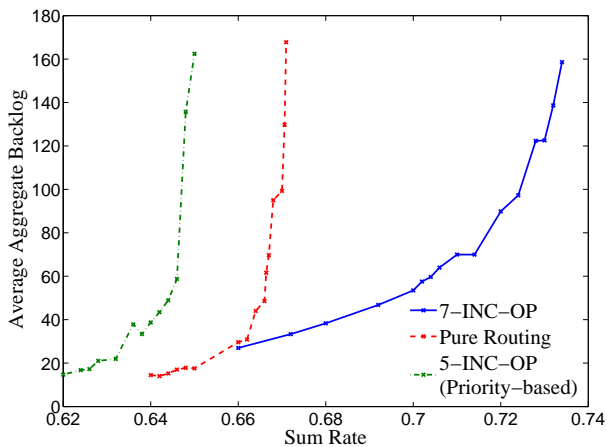}} \caption{\label{fig:overall}The backlog comparison with $\cq(t)$ chosen from $\{1,2,3,4\}$ and $\vec{p}^{(1)}=(0.14,0.06,0.56,0.24)$, $\vec{p}^{(2)}=(0.14,0.56,0.06,0.24)$, $\vec{p}^{(3)}=(0.04,0.16,0.16,0.64)$, and $\vec{p}^{(4)}=(0.49,0.21,0.21,0.09)$.\label{fig:overall-6}}
\end{figure}

\begin{figure}
	\centering
	\includegraphics[width=9cm]{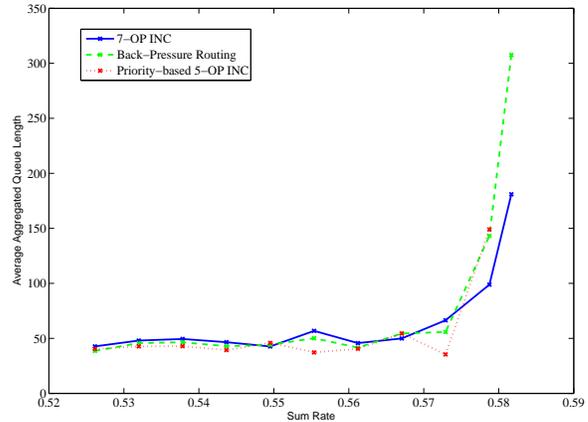}\\
	\caption{$(f_1,f_2,f_3,f_4)=(0.25,0.25,0.25,0.25)$ with $R_1=10\cdot R_2$. The backlog comparison with $\cq(t)$ chosen from $\{1,2,3,4\}$ and $\vec{p}^{(1)}=(0.14,0.06,0.56,0.24)$, $\vec{p}^{(2)}=(0.14,0.56,0.06,0.24)$, $\vec{p}^{(3)}=(0.04,0.16,0.16,0.64)$, and $\vec{p}^{(4)}=(0.49,0.21,0.21,0.09)$.}\label{fig:uneven_rate}
\end{figure}

\par Next we simulate the scenario of 4 different channel qualities: $\bigcq=\{1,2,3,4\}$. The varying channel qualities could model the situations like the different packet transmission rates and loss rates due to time-varying interference caused by the primary traffic in a cognitive radio environment. We assume four possible channel qualities with the corresponding probability distributions being $\vec{p}^{(1)}=(p^{(1)}_{\overline{d_1d_2}}, p^{(1)}_{d_1\overline{d_2}},p^{(1)}_{\overline{d_1}d_2},p^{(1)}_{d_1d_2})=(0.14,0.06,0.56,0.24)$, $\vec{p}^{(2)}=(0.14,0.56,0.06,0.24)$, $\vec{p}^{(3)}=(0.04,0.16,0.16,0.64)$, and $\vec{p}^{(4)}=(0.49,0.21,0.21,0.09)$ in both Figs.~\ref{fig:random} and \ref{fig:round}. The difference is that in Fig.~\ref{fig:random}, the channel quality $\cq(t)$ is i.i.d.\ with probability $(f_1,f_2,f_3,f_4)$ being $(0.15, 0.15, 0.35, 0.35)$. In Fig.~\ref{fig:round} the $\cq(t)$ is i.i.d.\ but with different frequency $(f_1,f_2,f_3,f_4)= (0.25, 0.25, 0.25, 0.25)$. \change{In Fig.~\ref{fig:periodic}, we consider the same set of channel qualities but choose $\cq(t)$ to be periodic with period 12 and the first period being $1, 1, 1, 2, 2, 2, 3, 3, 3, 4, 4, 4$. This scenario can be considered as an exmple of the Markovian channel quality.}  Again, we assume perfect fairness $(R_1,R_2)=(\theta,\theta)$. The sum-rate Shannon capacity is $R_1+R_2=0.716$ when $(f_1,f_2,f_3,f_4)=(0.15, 0.15, 0.35, 0.35)$ and $R_1+R_2=0.748$ when $(f_1,f_2,f_3,f_4)= (0.25, 0.25, 0.25, 0.25)$, and the pure routing sum-rate capacity is $R_1+R_2=0.625$ when $(f_1,f_2,f_3,f_4)=(0.15, 0.15, 0.35, 0.35)$ and $R_1+R_2=0.675$ when $(f_1,f_2,f_3,f_4)= (0.25, 0.25, 0.25, 0.25)$. We simulate our modified 7-OP INC, the priority-based solution in \cite{GeorgiadisTassiulas13}, and a standard back-pressure routing scheme \cite{tassiulas1992stability}.

\par Although the priority-based scheduling solution is provably optimal for fixed channel quality, it is less robust and can sometimes be substantially suboptimal (see Fig.~\ref{fig:round}) due to the ad-hoc nature of the priority-based policy. For example, as depicted by Figs.~\ref{fig:random} and \ref{fig:round}, the pure-routing solution outperforms the 5-operation scheme for one set of frequency $(f_1,f_2,f_3,f_4)$ while the order is reversed for another set of frequency. On the other hand, the proposed 7-operation scheme consistently outperforms all the existing solutions and has a stabiliby region matching the Shannon block-code capacity. We have tried many other combinations of time-varying channels. In all our simulations, the proposed DMW scheme always achieves the block-code capacity in \cite{WangHan14} and outperforms routing and any existing solutions \cite{AthanasiadouGeorgiadis13,GeorgiadisTassiulas13}. \change{Fig.~\ref{fig:uneven_rate} demonstrates the aggregated backlog result under a scenario that is similar to Fig.~\ref{fig:round} but with an extremely uneven arrival rate pair, $R_1=10\cdot R_2$. In this case, the network coding-based solution can barely provide significant throughput gain as proven in \cite{WangHan14}.}

\begin{table}
	\caption{Simulations for the settings in Figs.~\ref{fig:7vs5} and \ref{fig:overall-6} at different arrival rates. The representation ($x$; $y$) means that $x$ (resp.\ $y$) is for 90\% (resp.\ 95\%) of the optimal packet arrival rate.
		\label{tab:decoding_results}}
	\centering\small
	\begin{tabular}{c|c|c|c}
		&  \begin{tabular}{@{}c@{}}Avg.\ end-to- \\ end delay \\ (time slot)\end{tabular} & \begin{tabular}{@{}c@{}}Avg.\ receiver \\ buffer size \\ (no.\ of packets) \end{tabular}  & \begin{tabular}{@{}c@{}}Avg.\ receiver \\ buffer size of\\ existing solutions \end{tabular}  \\ \hline
		Fig.~\ref{fig:7vs5}	&  (34,84; 85.20) & (7.35; 17.94) & (28.74; 128.20)  \\ \hline
		Fig.~\ref{fig:random}	& (33.41; 72.03) & (5.19; 12.45) & (14.20; 39.34)  \\ \hline
		Fig.~\ref{fig:round}	& (37.56; 75.23) & (6.20; 13.62) & (16.94; 41.66)
	\end{tabular}
\end{table}

\par Using the same settings as in Figs.~\ref{fig:7vs5}, \ref{fig:random}, and \ref{fig:round}, Table~\ref{tab:decoding_results} examines the corresponding end-to-end delay and buffer usage. Specifically the end-to-end delay measures the time slots each packet takes from its arrival at $s$ to the time slot it is successfully decoded by its intended destination, which includes the queueing, propagation, and decoding delay. The buffer size is measured at the receivers according to our buffer management policy in Section~\ref{subsec:buf_mang_rx}.   The statistics are derived under either 90\% or 95\% of the optimal sum arrival rate, which corresponds to 0.9 or 0.95 packets/slot; 0.64 or 0.68 packets/slot; and 0.67 or 0.71 packets/slot for the settings in Figs.~\ref{fig:7vs5}, \ref{fig:random}, and \ref{fig:round}, respectively. The last column of Table~\ref{tab:decoding_results} also reports the buffer usage if we use the existing $(i^*,j^*)$-based buffer pruning policy described in Section~\ref{subsec:buf_mang_rx}.

We notice that our scheme has small delay and buffer usage at 90\% of the optimal arrival rate and the delay and  buffer size are still quite manageable even at 95\% of the optimal arrival rate. It is worth noting that 4 out of the 6 chosen arrival rates are beyond the stability region of the best existing routing/NC solutions \cite{GeorgiadisTassiulas13,AthanasiadouGeorgiadis13,Paschos12} and those schemes will thus have exploding delay and buffer sizes under those cases. Table~\ref{tab:decoding_results} also confirms that the solution proposed in Section~\ref{subsec:buf_mang_rx} can significantly reduce the buffer size of the existing NC solutions \cite{KattiRahulHuKatabiMedardCrowcroft06,GeorgiadisTassiulas13,AthanasiadouGeorgiadis13,Paschos12}.

\begin{figure}
\centering
  \includegraphics[width=9cm]{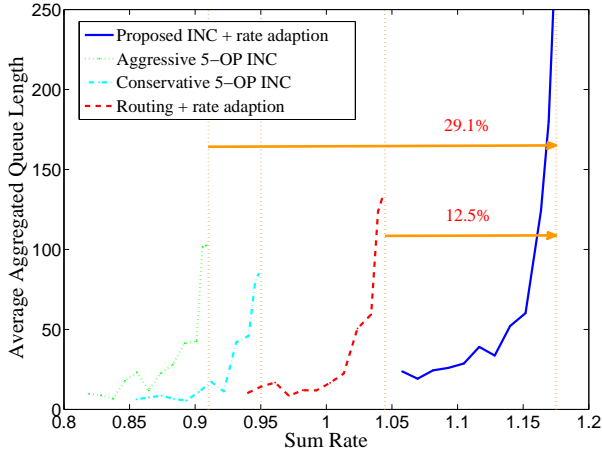}\\
  \caption{The backlog of four different schemes for rate adaptation with two possible (error-correcting-code rate,modulation) combinations. The back-pressure-based INC scheme in \cite{AthanasiadouGeorgiadis13} is used in both aggressive and conservative 5-OP INC, where the former always chooses the high-throughput (rate,modulation) combination while the latter always chooses the low-throughput (rate,modulation) combination.}\label{fig:acm}
\end{figure}

\par Our solution in Section~\ref{subsec:extention_ACM} is the first dynamic INC design that achieves the optimal linear INC capacity with rate-adaptation \cite{WangHan14}. Fig.~\ref{fig:acm} compares its performance with existing routing-based rate-adaptation scheme and the existing INC schemes, the latter of which are designed without rate adaptation. We assume there are two available (error-correcting-code rate,modulation) combinations; and the first (resp.\ second) combination takes 1 second (resp.\ $\frac{1}{3}$ second) to finish transmitting a single packet. I.e., the transmission rate of the second combination is 3 times faster.

\par We further assume the packet delivery probability is $\vec{p}=(p_{\overline{d_1d_2}}, p_{d_1\overline{d_2}},p_{\overline{d_1}d_2},p_{d_1d_2})=(0.005,0.095,0.045,0.855)$ if the first combination is selected and $\vec{p}=(0.48,0.32,0.12,0.08)$ for the second combination. That is, the low-throughput combination is likely to be overheard by both destinations and the high-throughput combination has a much lower success probability. We can compute the corresponding Shannon capacity by modifying the equations in \cite{WangHan14}. We then use the proportional fairness objective function $\xi(R_1,R_2)=\log(R_1)+\log(R_2)$ and find the maximizing  $R_1^*$ and $R_2^*$ over the Shannon capacity region, which are $(R_1^*, R_2^*)=( 0.6508, 0.5245)$  packets per second.

After computing $(R_1^*, R_2^*)$, we assume the following dynamic packet arrivals. We define $(R_1,R_2)=\theta\cdot (R_1^*,R_2^*)$ for any given $\theta\in(0,1)$. For any experiment (i.e., for any given $\theta$), the arrivals of session-$i$ packets is a Poisson random process with rate $R_i$ packets per second for $i=1,2$.

\par Each point of the curves of Fig.~\ref{fig:acm} consists of 10 trials and each trial lasts for $10^5$ seconds. We compare the performance of our scheme in Section~\ref{subsec:extention_ACM} with (i) Pure-routing with rate-adaptation; (ii) aggressive 5-OP INC, i.e., use the scheme in \cite{AthanasiadouGeorgiadis13} and always choose combination 2; and (iii) conservative 5-OP INC, i.e., use the scheme in \cite{AthanasiadouGeorgiadis13} and always choose combination 1. We also plot the optimal routing-based rate-adaptation rate and the optimal Shannon-block-code capacity rate as vertical lines.

\par Since our proposed scheme jointly decides which (rate,modulation) combination and which INC operation to use in an optimal way, see \eqref{eq:new-sch-ACM}, the stability region of our scheme matches the Shannon capacity with rate-adaptation. It provides $12.51\%$ throughput improvement over the purely routing-based rate-adaptation solution, see Fig.~\ref{fig:acm}. 

\par Furthermore, if we perform INC but always choose the low-throughput (rate,modulation), as suggested in some existing works \cite{RayanchuSenWuBanerjeeSengupta08}, then the largest sum-rate $R_1+R_2=\theta_\text{cnsv.\ 5-OP}^*(R_1^*+R_2^*)=0.9503$, which is worse than pure routing with rate-adaptation $\theta_\text{routing,RA}^*(R_1^*+R_2^*)=1.0446$. Even if we always choose the high-throughput (rate,modulation) with 5-OP INC, then the largest sum-rate $R_1+R_2=\theta_\text{aggr.\ 5-OP}^*(R_1^*+R_2^*)=0.9102$ is even worse than the conservative 5-OP INC capacity. We have tried many other rate-adaptation scenarios. In all our simulations, the proposed DMW scheme always achieves the capacity and outperforms pure-routing, conservative 5-OP INC, and aggressive 5-OP INC.

\par It is worth emphasizing that in our simulation, for any fixed (rate,modulation) combination, the channel quality is also fixed. Therefore since 5-OP scheme is throughput optimal for fixed channel quality \cite{GeorgiadisTassiulas09}, it is guaranteed that the 5-OP scheme is throughput optimal when using a fixed (rate,modulation) combination. Our results thus show that using a fixed (rate,modulation) combination is the main reason of the suboptimal performance. At the same time, the proposed scheme in \eqref{eq:DMW-schedule}, \eqref{eq:DMW-update-new}, and \eqref{eq:new-sch-ACM} can dynamically decide which (rate,modulation) combination to use for each transmission and achieve the largest possible stability region.

\section{Conclusion\label{sec:conclusion}}

We have proposed a new 7-operation INC scheme together with the corresponding scheduling algorithm to achieve the optimal downlink throughput of the 2-flow access point network with time varying channels. Based on binary XOR operations, the proposed solution  admits ultra-low encoding/decoding complexity with efficient buffer management and minimal communication and control overhead. The proposed algorithm has also been generalized for rate adaptation and it again robustly achieves the optimal throughput in all the numerical experiments. A byproduct of this paper is a throughput-optimal scheduling solution for SPNs with random departure, which could further broaden the applications of SPNs to other real-world applications.

\bibliographystyle{IEEEtran}
\bibliography{WCK}

\end{document}